\documentclass[final,twocolumn]{IEEEtran}
\usepackage{amsmath}
\usepackage{amssymb}
\usepackage{array}
\usepackage{fixmath}
\usepackage{graphicx}
\usepackage{cite}
\usepackage{color}
\usepackage{changes}
\usepackage{algorithm}
\usepackage{algorithmic}
\usepackage{acronym}
\usepackage{subcaption}
\usepackage{bm}
\usepackage{lipsum}
\usepackage{mathtools}
\usepackage{cuted}
\usepackage{multirow}
\usepackage{multicol}

\graphicspath{{figures/}} 

\DeclareMathAlphabet{\mathbit}{OML}{cmr}{bx}{it}

\newacro{TDD}{time-division duplexing}
\newacro{CSI}{channel state information}
\newacro{DL}{downlink}
\newacro{UL}{uplink}
\newacro{BS}{base station}
\newacro{MS}{mobile station}
\newacro{MSE}{mean square error}
\newacro{MMSE}{minimum mean square error}
\newacro{SVD}{singular value decomposition}
\newacro{AM}{alternating minimization}
\newacro{OFDM}{orthogonal frequency-division multiplexing}
\newacro{mmWave}{millimeter wave}
\newacro{OMP}{orthogonal matching pursuit}
\newacro{MIMO}{multiple-input multiple-output}
\newacro{RF}{radio-frequency}
\newacro{LS}{least squares}
\newacro{MRC}{maximum ratio combiner}
\newacro{ZF}{zero-forcing}
\newacro{CS}{compressive sensing}
\newacro{ULA}{uniform linear array}
\newacro{ADC}{analog-to-digital converter}
\newacro{AoA}{angles-of-arrival}
\newacro{AoD}{angles-of-departure}
\newacro{CRLB}{Cram\'{e}r-Rao lower bound} 
\newacro{NMSE}{normalized mean squared error}
\newacro{CB}{coordinated beamforming}
\newacro{NMSE}{normalized mean-squared error}
\newacro{SINR}{signal-to-interference-plus-noise ratio}
\newacro{LLF}{log-likelihood function}
\newacro{SW-OMP}{simultaneous weighted - orthogonal matching pursuit}
\newacro{SS-SW-OMP+Th}{subcarrier-selection simultaneous weighted - orthogonal matching pursuit + thresholding}

\DeclareMathOperator{\vect}{vec}

\newcounter{MYtempeqncnt}

\newcommand{\SmBlc}{{\mathbf{S}}_{m}}

\newcommand{\td}{\mathrm{td}}

\newcommand{\Ts}{T_\mathrm{s}}
\newcommand{\SBlc}{{\mathbf{S}}_{1}}

\newcommand{\Arx}{{\mathbf{A}}_{\mathrm{R}}}
\newcommand{\Atx}{{\mathbf{A}}_{\mathrm{T}}}
\newcommand{\Atxbar}{\bar{\mathbf{A}}_{\mathrm{T}}}
\newcommand{\at}{{\mathbf{a}}_{\mathrm{T}}}
\newcommand{\ar}{{\mathbf{a}}_{\mathrm{R}}}
\newcommand{\Gc}{G_{\mathrm{c}}}

\def\hc{\bh_{\mathrm{c}}}

\newcommand{\Gr}{{G_\text{RX}}}
\newcommand{\Gt}{{G_\text{TX}}}

\newcommand{\NTX}{{N_\text{T}}}
\newcommand{\NRX}{{N_\text{R}}}

\def\bee{{\mathbf{e}}}

\def\bh{{\mathbf{h}}}

\def\bo{{\mathbf{o}}}
\def\bp{{\mathbf{p}}}

\def\bs{{\mathbf{s}}}

\def\bx{{\mathbf{x}}}
\def\by{{\mathbf{y}}}

\def\b0{{\mathbf{0}}}

\def\bF{{\mathbf{F}}}

\def\bH{{\mathbf{H}}}
\def\bI{{\mathbf{I}}}

\def\bR{{\mathbf{R}}}
\def\bS{{\mathbf{S}}}

\def\bW{{\mathbf{W}}}

\def\sf0{{\mathsf{0}}}

\def\bsf0{{\bm{\mathsf{0}}}}

\def\bbC{{\mathbb{C}}}

\def\bbR{{\mathbb{R}}}

\def\bphi{{\bf \Phi}}
\def\bpsi{{\bf \Psi}}

\begin{document}

\title{
Multidimensional orthogonal matching pursuit: theory and application to high accuracy joint localization and communication at mmWave}
\author{\IEEEauthorblockN{Joan Palacios,~Nuria Gonz\'alez-Prelcic and Cristian Rusu}\\
	\IEEEauthorblockA{%
		North Carolina State University, USA\\
		Email:\{\texttt{jbeltra,ngprelcic}\}\texttt{@ncsu.edu}}
}

\maketitle

\begin{abstract}
Greedy approaches in general, and orthogonal matching pursuit in particular, are the most commonly used sparse recovery techniques in a wide range of applications. The complexity of these approaches is highly dependent on the size of the dictionary chosen to represent the sparse signal. When the dictionary has to be large to enable high accuracy reconstructions, greedy strategies might however incur in prohibitive complexity. In this paper, we propose first the formulation of  a new type of sparse recovery problems where the sparse signal is represented by a set of independent and smaller dictionaries instead of a large single one. Then, we derive a low complexity multdimensional orthogonal matching pursuit (MOMP) strategy for sparse recovery with a multdimensional dictionary. The projection step is performed iteratively on every dimension of the dictionary while fixing all other dimensions to achieve high accuracy estimation at a reasonable complexity. 
Finally, we formulate the problem of high resolution time domain channel estimation at millimeter wave (mmWave) frequencies  as a multidimensional sparse recovery problem that can be solved with MOMP. The channel estimates are later transformed into high accuracy user position estimates exploiting a new localization algorithm that leverages the particular geometry of indoor channels. Simulation results show the effectiveness of MOMP for high accuracy localization at millimeter wave frequencies when operating in realistic 3D scenarios, with practical MIMO architectures feasible at mmWave, and without resorting to perfect synchronization assumptions that simplify the problem. 
\end{abstract}

\section{Introduction}

Sparse recovery algorithms provide computationally efficient ways to reconstruct sparse signals from a number of linear observations smaller than the dimension of the signal.
Applications  include signal approximation, compressive imaging, channel estimation, or spectrum sensing to name a few. 
The basic premise for sparse recovery is that the signal $\bx \in \bbR^n$  to be recovered is sparse, i.e. can be represented in terms of some basis or dictionary $\bpsi \in \bbR^{n\times n}$ as $\bx=\bpsi\bs$, with a number of non zero coefficientes in $\bs$ much smalller than $n$. 
The problem of reconstructing this sparse signal can be formulated as a minimization of the number of non zero coefficients ($\ell_0$-norm) in $\bs$ subject to an observation $\bo \in \bbR^m$, with $m<n$. To compute the observation we assume that a linear measurement matrix denoted as $\bphi \in \bbR^{m\times n}$ is applied, i.e. $\bo=\bphi\bx$. With these definitions, the sparse reconstruction problem becomes
\begin{equation}
\min_s \| \bs \|_0 \quad \text{s.t}. \quad  \bo=\bphi \bpsi \bs.
\end{equation}
Since this problem is NP-hard, an optimization based on the $\ell_1$-norm was proposed instead, which brings down  the problem to the basis pursuit optimization \cite{Candes2006,Chen1998}.

To reduce computational cost of the methods used to solve the $\ell_1$-norm relaxation, greedy approaches were proposed \cite{Tropp2004}. While both relaxation and greedy techniques have similar theoretical sparse recovery guarantees~\cite{TROPP2006572,TROPP2006589} and similar polynomial numerical complexity, in practice, they produce different sparse solutions, and their running times heavily depends on many practical implementation factors. Greedy algorithms are, in general, the preferred sparse recovery technique when low running time is desired, because of their ability to efficiently solve for multiple target vectors $\by$, and because they exploit modern computing architectures. In particular, the Orthogonal Matching Pursuit (OMP)~\cite{10.1117/12.173207} algorithm enjoys efficient software implementations that exploit blocking and batch of the dictionaries and target vectors, and efficient updating techniques for Cholesky factorizations to achieve fast sparse approximations with a low memory footprint~\cite{rubinstein2008efficient,6692175}. 

Greedy sparse recovery has also been used to estimate the millimeter wave MIMO channel matrix, which can be modeled as a sparse vector after vectorization \cite{mmWavetutorial}.  OMP is the most common sparse recovery method used in prior work on mmWave channel estimation  \cite{lee2014exploiting,Venugopal2017, SWOMP2018, Coma2018, Wu2019, Zhu2019}.  Different works describe approaches in the time domain \cite{Venugopal2017,Zhu2019}, in the frequency domain \cite{lee2014exploiting,Wu2019,SWOMP2018, Coma2018}, assuming a narrowband channel model \cite{lee2014exploiting,Wu2019}, or considering a frequency selective one \cite{Venugopal2017, SWOMP2018,Zhu2019,Coma2018}. In general, for the frequency domain approaches, the dictionaries are built as a Kronecker product of the array steering vectors at the transmitter and at the receiver evaluated on a grid for the angle of departure (AoD) and the angle of arrival (AoA). In the time domain approaches, an additional component is needed to represent the delay domain  when building the dictionary as a Kronecker product \cite{Venugopal2017}.
Although some greedy methods that exploit the structure of Kronecker dictionaries to reduce complexity have been proposed and in other signal processing applications \cite{6288476,10.1162/NECO_a_00385,7520367,10.1162/neco_a_01304}, unfortunately, none of them can be applied to the mmWave channel estimation problem. This is due to the fact that the equivalent measurement matrix that appears in the compressive channel recovery problem includes the combined effect of the described Kronecker dictionary and the sensing matrix, based on the training precoders and combiners used at the transmitter and the receiver to sound the channel. This way, the Kronecker structure disappears in the equivalent measurement matrix $\bphi \bpsi$, and sparse recovery based on conventional OMP implementations remains the preferred option for channel estimation at mmWave.

Channel estimation is one of the key operations in a mmWave receiver to enable configuration of the antenna arrays and link establishment. However, the capabilities that MIMO technology can offer at mmWave bands go beyond high data rate communication.
Thus, the angular and delay resolvability that can be achieved  with mmWave MIMO enables the usage of the output of the channel estimation algorithm for high accuracy device localization as a byproduct of communication. Once the channel is estimated, the geometric relationships between the channel parameters (AoA/AoD and delays)  and the position and orientation of the transmitter and the receiver  can be exploited to obtain position information as in \cite{Shahmansoori2018, Wymeersch2018, Talvitie2019, Jiang2021}, without need of establishing links with more than an access point/base station to exploit triangulation \cite{palacios2019leap}.  Achieving good localization performance, though, requires very high resolution channel estimates, both for the angular parameters and the delays of every channel path. 

State of the art algorithms for compressive channel estimation at mmWave cannot provide this required accuracy in realistic settings. 
For example, works on millimeter wave channel estimation based on OMP report interesting results when applied to systems that operate with moderately large linear arrays and dictionaries \cite{lee2014exploiting,Wu2019,SWOMP2018, Coma2018}. 
Operation with realistic antennas to enable 3D localization, like large planar arrays, pose serious complexity and memory challenges to OMP-based mmWave channel estimation approaches. Furthermore, when additionally considering the stringent resolution specifications in the grids used to build the dictionary, required to enable high accuracy angular and delay estimates needed for 3D localization, state of the art approaches simply cannot run in a reasonable time, and/or require a memory size not available in current devices. 
To enable the evaluation of localization capabilities, prior work on joint compressive channel estimation and localization  resorts to some impractical assumptions. For example, many works assume uniform linear arrays \cite{Shahmansoori2018,Talvitie2019} or small size planar arrays \cite{Jiang2021}, a non bandlimited channel model that enhances sparsity \cite{Shahmansoori2018,Wymeersch2018,Jiang2021}, or a fully digital MIMO architecture at the device unfeasible at mmWave  \cite{Shahmansoori2018,Talvitie2019,Jiang2021},  which simplifies the localization scenario and the compressive channel estimation problem to be solved. To achieve high localization accuracy in practical settings, precise channel estimators are needed that do not suffer from high complexity encountered when operating in a realistic 3D channel with planar arrays and a hybrid or analog architecture at the device. Finally, an additional limitation of previous work on joint localization and channel estimation at mmWave,  not related to computational complexity, is the perfect synchronization assumption. Thus, many works also assume that the the transmitter and receiver are being triggered at the same time, so that the absolute delay of the LoS component can be obtained during the channel estimation process \cite{Shahmansoori2018,Wymeersch2018,Talvitie2019,Jiang2021}. 
To improve realizability of the algorithms, these impractical synchronization assumption also has to be discarded. 

\subsection{Contributions}
In this paper, we present several contributions to overcome complexity limitations of greedy sparse recovery algorithms in general, and when applied to the problem of channel estimation and localization at mmWave.  

First, unlike prior work on greedy sparse recovery, we consider the possibility of operating with multiple independent dictionaries, instead of a single dictionary that can become prohibitively large in some applications. We formulate first the corresponding sparse recovery problem with a multidimensional dictionary built as a product of independent dictionaries, and then propose a fast algorithm to solve it, named multidimensional orthogonal matching pursuit (MOMP). In formulating the algorithm, we account for the practical linear measurement matrix that needs to be used in conjunction with the dictionary for sparse recovery. Complexity and memory requirements of the proposed algorithm 
are analyzed and compared to those of conventional greedy sparse recovery approaches.

Then, we formulate the channel estimation problem at mmWave bands as a sparse recovery algorithm with mutiple sparsifying dictionaries that represent the channel features in the angular and delay domains, building upon our preliminary work in \cite{Palacios2022Eusipco}. To overcome limitations of previous work, the proposed formulation considers a realistic band-limited channel model including filtering effects at the transmitter and the receiver, and a 3D scenario where operation with large planar arrays instead of linear is a must. In addition, the channel measurements considered in our formulation are taken using a practical hybrid or analog MIMO architecture at the devices, and a hybrid architecture at the access point. The channel estimation problem is then solved by exploiting the proposed MOMP algorithm. Unlike in 
\cite{Palacios2022Eusipco}, we consider users that can operate with either an analog MIMO architecture or a hybrid architecture.

\begin{figure*}[t!]
\centerline{\includegraphics[width=0.9\linewidth, trim={50px 100px 55px 60px}, clip]{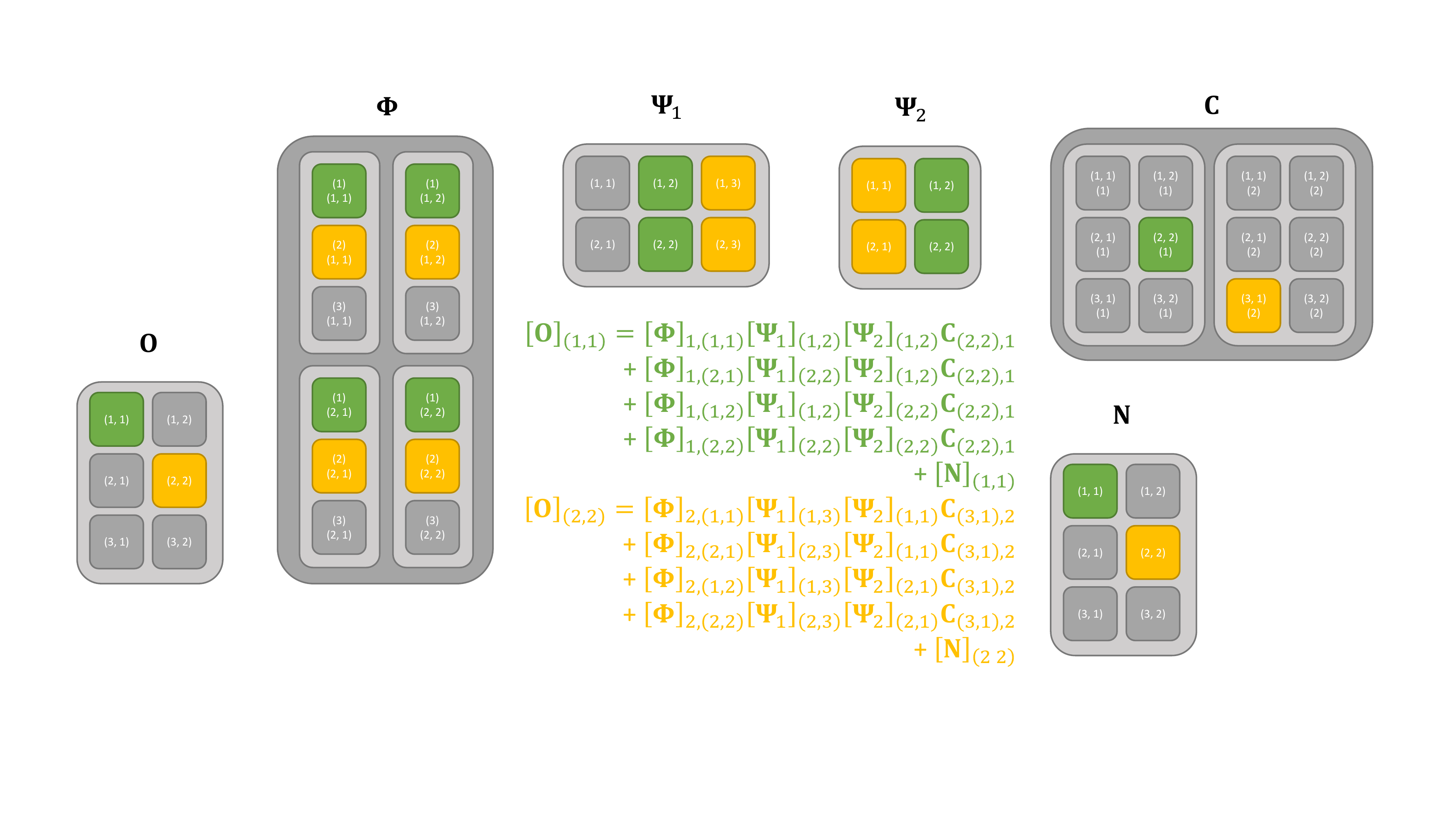}}
\caption{Eq~\ref{eq:MP_multi} entries calculation example when $\mathcal{C} = \{[2, 2], [3, 1]\}$.}
\label{fig:meas_ex}
\end{figure*}

Finally, we develop a localization method for indoor scenarios that estimates the position of the device by exploiting the 
AoA, AoD and time difference of arrival (TDoA), i.e. relative delays, for each estimated channel path. No perfect synchronization between the transmitter and the receiver needs to be assumed. The derived geometric relationships between the channel path parameters and the position of the device and the access point leverage the fact that in an indoor scenario the first order reflections in the channel are created by either vertical walls or horizontal surfaces that model the floor or the ceiling.

Simulation results over a channel data set generated by ray tracing show how high accuracy localization and channel estimation can be enabled in a realistic environment by exploiting independent dictionaries in the angular and delay domains and MOMP, while conventional greedy approaches cannot even run due to complexity and/or memory limitations when operating with large planar arrays and practical simulation conditions. To the best of out knowledge, there is no prior work on greedy sparse recovery with multidimensional dictionaries. Furthermore, there are not alternative solutions to joint localization and compressive channel estimation at millimeter wave that can operate in a 3D scenario,  with the realistic channel model and MIMO architectures that we assume in our work, without the perfect synchronization assumption, and that have been experimentally evaluated in a realistic ray tracing model of an indoor scenario.


\subsection{Notation}
We use the following notation throughout the paper.
$x$, ${\bf x}$, ${\bf X}$ and $\mathcal{X}$ will be the styles for scalar, vector, matrix or tensor and set.
Regarding sub/supper-indexes, $x$ and ${\rm x}$ are used to denote scalar and categorical values respectively.
$[{\bf x}]_n$ denotes the $n$-th entry of ${\bf x}$.
For a 2D matrix ${\bf X}$, $[{\bf X}]_{a, b}$, $[{\bf X}]_{a, :}$ and $[{\bf X}]_{:, b}$ are respectively, the element in the $a$-th row and $b$-th column, the $a$-th row and the $b$-th column, this notation is extended to the case of tensors with multi-indexes acting like multiple indexes, such as $[{\bf X}]_{{\bf a}, b} = [{\bf X}]_{a_1, a_2, b}$ for ${\bf a} = [a_1, a_2]$.
We use the operator $\|{\bf x}\|$, $\|{\bf X}\|$ to denote the Euclidean and Frobenius norms of ${\bf x}$ and ${\bf X}$ respectively.
We do not consider $0$ to be a natural number, therefore we start enumerating from 1.
Regarding matrix multiplication, we consider all one dimensional vectors to be  column vectors.

\section{Sparse recovery via multidimensional orthogonal matching pursuit}\label{sec:MOMP}

\subsection{Background}
The problem of sparse recovery with multiple measurements can be written in terms of an observation matrix that contains a set of measurements of the sparse signal to be recovered, a sparsifying dictionary that enables the representation of the signal of interest with a few large coefficients, and a sensing matrix that represents the linear measurement process.
Let ${\bf O}\in\mathbb{C}^{N^{\rm q}\times N^{\rm m}}$ be the observation matrix, which contains a set of $N^{\rm m}$ observations, each one of dimension $N^{\rm q}$. 
The sparsifying dictionary is denoted by the matrix ${\bf \Psi}\in\mathbb{C}^{N^{\rm s}\times N^{\rm a}}$, where $N^{\rm a}$ is the number of atoms in the dictionary and $N^{\rm s}$ is the size of each atom. The coefficients of the signal in the dictionary ${\bf \Psi}$  are represented by the sparse matrix ${\bf C}\in\mathbb{C}^{N^{\rm a}\times N^{\rm m}}$. Finally, the measurement matrix is denoted as ${\bf \Phi}\in\mathbb{C}^{N^{\rm q}\times N^{\rm s}}$. With these definitions, the sparse recovery problem can be formulated as 
\begin{equation}\label{eq:MP_single}
\min_{\bf C}\left(\|{\bf O}-{\bf \Phi}{\bf \Psi}{\bf C}\|^2\right).
\end{equation}
To be coherent in notation with previous works, each of the columns of ${\bf O}$ is called an observation, each column of ${\bf \Psi}$ is called an atom, and the set $\mathcal{C}$ of indexes $i\leq N^{\rm a}$ such that $\|{\bf C}_{i, :}\|>0$ is called the support. 
The sparsity condition applied to ${\bf C}$ is $|\mathcal{C}| \leq N_{\rm p} <<  N^{\rm a}$.
We will  consider the set $\mathcal{C}$ as a sorted set,  so that its elements are associated with a set of natural indexes, i.e. $\mathcal{C} = \{c_1, c_2, \ldots, c_{N_{\rm p}}\}$, where $N_{\rm p}$ denotes the sparsity level.
\begin{figure*}[t!]
\centerline{\includegraphics[width=0.6\linewidth, trim={185px 100px 260px 100px}, clip]{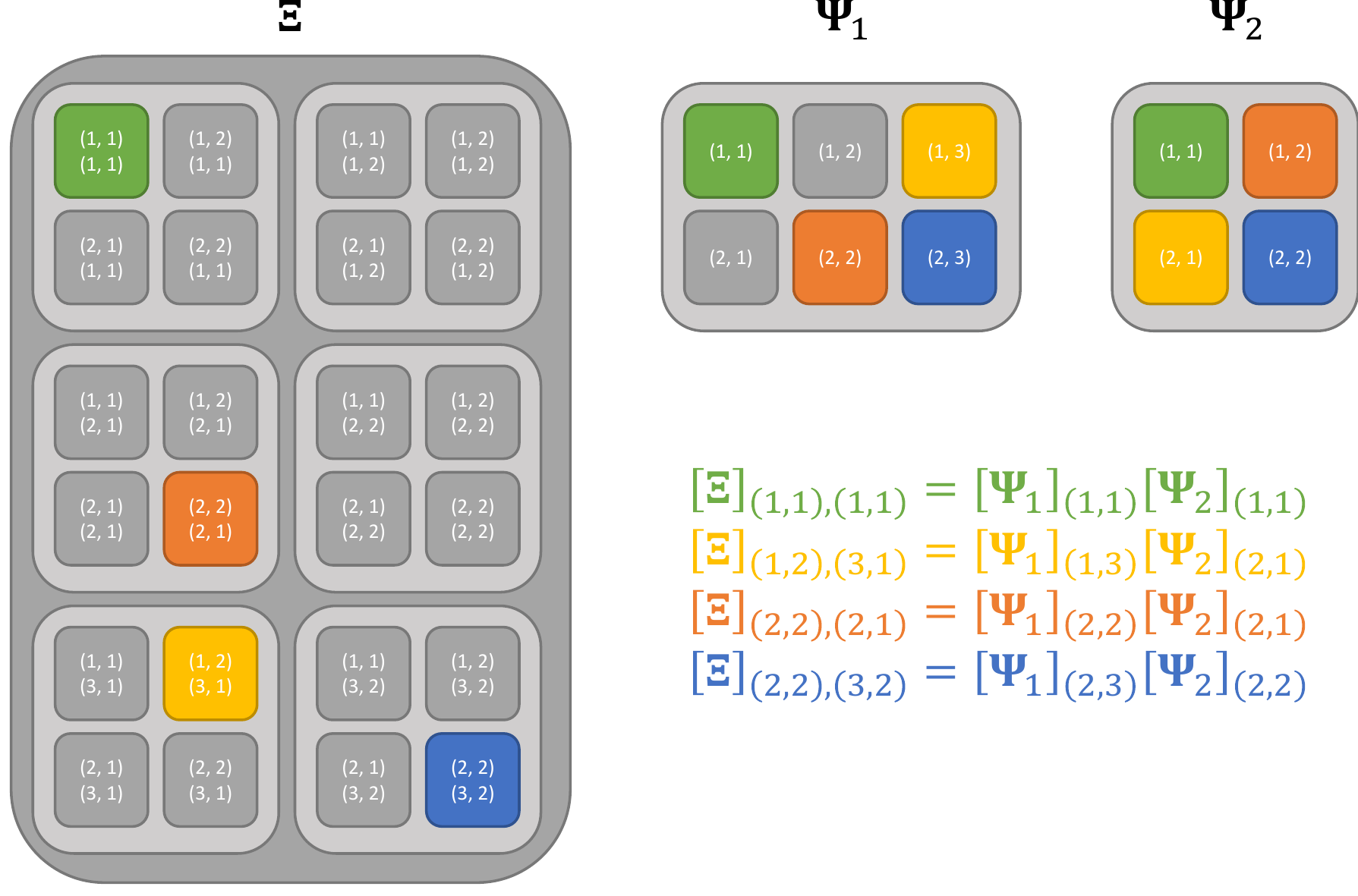}}
\caption{Entries calculation example for the matrix ${\bf \Xi}$.}
\label{fig:xi_ex}
\end{figure*}
\subsection{Sparse recovery with a multidimensional dictionary}
Let us  assume we have $N_{\rm D}$ dictionaries, with the $k$-th dictionary ${\bf \Psi}_{k}\in\mathbb{C}^{N_k^{\rm s}\times N_k^{\rm a}}$ consisting of $N_k^{\rm a}$ atoms in $\mathbb{C}^{N_k^{\rm s}}$.
We also define the coefficients of the sparse signal in the set of dictionaries as ${\bf C}\in\mathbb{C}^{N_{1}^{\rm a}\times\ldots\times N_{N_{\rm D}}^{\rm a}\times N^{\rm m}}$. In the multidimensional case, the measurement matrix can be written as a tensor, i.e.,  ${\bf \Phi}\in\mathbb{C}^{N^{\rm q}\times N_{1}^{\rm s}\times\ldots\times N_{N_{\rm D}}^{\rm s}}$.
We define the set of entry coordinate combinations $\mathcal{I} = \{{\bf i}=(i_1, \ldots, i_{N_{\rm D}})\in\mathbb{N}^{N_{\rm D}}\text{ s.t. }i_k \leq N_k^{\rm s}\quad\forall k \leq N_{\rm D}\}$, and the set of dictionary index combinations $\mathcal{J} = \{{\bf j}=(j_1, \ldots, j_{N_{\rm D}})\in\mathbb{N}^{N_{\rm D}}\text{ s.t. }j_{d} \leq N_k^{\rm a}\quad\forall k \leq N_{\rm D}\}$ to cycle over each dictionary atom entry index and dictionary atom index, respectively.
The equivalent multidimensional sparse recovery problem is now defined as
\begin{equation}\label{eq:MP_multi}
\min_{\bf C}\left(\sum_{i_{\rm m} = 1}^{N^{\rm m}}\left\|[{\bf O}]_{:, i_{\rm m}}\hspace*{-1mm}-\hspace*{-1mm}\sum_{{\bf i}\in\mathcal{I}}\sum_{{\bf j}\in\mathcal{J}}[{\bf \Phi}]_{:, {\bf i}}\left(\prod_{k = 1}^{N_{\rm D}}[{\bf \Psi}_{k}]_{i_k, j_k}\right)\hspace*{-1mm}[{\bf C}]_{{\bf j}, i_{\rm m}}\right\|^2\right)
\end{equation}
and an example of the interaction of the different elements in \eqref{eq:MP_multi}  is depicted in Fig.~\ref{fig:meas_ex}.

A way to interpret \eqref{eq:MP_multi} is to see the double multi-index sum
\[\sum_{{\bf i}\in\mathcal{I}}\sum_{{\bf j}\in\mathcal{J}}[{\bf \Phi}]_{:, {\bf i}}\left(\prod_{k = 1}^{N_{\rm D}}[{\bf \Psi}_{k}]_{i_k, j_k}\right)[{\bf C}]_{{\bf j}, i_{\rm m}}\]
as a two matrices and one vector multiplication.
This is easier to see when substituting the middle term by elements of the tensor ${\bf \Xi}\in\mathbb{C}^{N_{1}^{\rm s}\times\ldots\times N_{N_{\rm D}}^{\rm s}\times N_{1}^{\rm a}\times\ldots\times N_{N_{\rm D}}^{\rm a}}$, defined as
\[[{\bf \Xi}]_{{\bf i}, {\bf j}} = \prod_{k = 1}^{N_{\rm D}}[{\bf \Psi}_{k}]_{i_k, j_k}.\]
Under this definition, the double multi-index sum can be written as
\begin{equation}\label{eq:extended_prod_x}
\sum_{{\bf i}\in\mathcal{I}}\sum_{{\bf j}\in\mathcal{J}}[{\bf \Phi}]_{:, {\bf i}}[{\bf \Xi}]_{{\bf i}, {\bf j}}[{\bf C}]_{{\bf j}, i_{\rm m}},
\end{equation}
which resembles the two matrices and one vector multiplication formula ${\bf A}{\bf B}{\bf c} = \sum_i\sum_j[{\bf A}]_{:, i}[{\bf B}]_{i, j}[{\bf c}]_{j}$ when the multi-index is linearized.
An illustration of how ${\bf \Xi}$ is built can be found in Fig.~\ref{fig:xi_ex}.
For this problem, the support is defined as the set $\mathcal{C}$ of tuples ${\bf j}\in\mathcal{J}$ such that $\|[{\bf C}]_{{\bf j}, :}\| > 0$.

To prove that the proposed multidimensional sparse recovery problem is an extension of the original formulation in \eqref{eq:MP_single}, we simply set $N_{\rm D} = 1$ to get to the expression
\begin{equation}
\min_{\bf C}\left(\|{\bf O}-{\bf \Phi}{\bf \Psi}_1{\bf C}\|^2\right),
\end{equation}
which is the same one as for the original problem in \eqref{eq:MP_single} when dropping the sub-index.
Furthermore, this problem is equivalent to the classical one when considering $\overline{\bf \Phi}$, $\overline{\bf \Psi}$ and $\overline{\bf C}$ as the ${\bf \Phi}$, ${\bf \Psi}$ and ${\bf C}$ variables in the classical formulation, for $\overline{\bf \Phi}\in\mathbb{C}^{N^{\rm q}\times\left(\prod_{k=1}^{N_{\rm D}}N_k^{\rm s}\right)}$ being the reshaped version of ${\bf \Phi}$, $\overline{\bf \Psi} = \bigotimes_{d=1}^{N_{\rm D}}{\bf \Psi}_{k}$, or equivalently the reshaped version of ${\bf \Xi}$, and $\overline{\bf C}\in\mathbb{C}^{\left(\prod_{k=1}^{N_{\rm D}}N_k^{\rm a}\right)\times N^{\rm m}}$ being the reshaped version of ${\bf C}$.
This is due the fact that since $\overline{\bf \Phi}$, $\overline{\bf \Psi}$ and $\overline{\bf C}$ are reshaped versions of ${\bf \Phi}$, ${\bf \Xi}$ and ${\bf C}$, the expression \eqref{eq:extended_prod_x} can be written as $[\overline{\bf \Phi}\overline{\bf \Psi}\overline{\bf C}]_{:, i_{\rm m}}$, such that \eqref{eq:MP_multi} becomes
\begin{equation}
\min_{\overline{\bf C}}\left(\sum_{i_{\rm m} = 1}^{N^{\rm m}}\left\|[{\bf O}]_{:, i_{\rm m}}-[\overline{\bf \Phi}\overline{\bf \Psi}\overline{\bf C}]_{:, i_{\rm m}}\right\|^2\right),
\end{equation}
which is equivalent to \eqref{eq:MP_single}. This is another way to prove that the proposed multidimensional sparse recovery problem is an extension of the original formulation with a single dictionary if the multiple dictionaries are combined through Kronecker product.
The definition of the support for the multidimensional problem is consistent with the equivalent classical problem when considering $\overline{\bf C}$.

\begin{figure*}[!t]
\normalsize
\setcounter{MYtempeqncnt}{\value{equation}}
\setcounter{equation}{9}
\begin{equation}
\max_{j_k}\left(\frac{\|\sum_{{\bf i}\in\mathcal{I}}[{\bf O}_{\rm \Phi}]_{:, {\bf i}}[{\bf \Psi}_{k}]_{i_k, \hat{j}_k}\prod_{k'\in\mathcal{E}}[{\bf \Psi}_{k'}]_{i_{k'}, \hat{j}_{k'}}\prod_{k'\in\hat{\mathcal{E}}}[{\bf \Psi}_{k'}]_{i_{k'}, \hat{j}_{k'}}\|^2}{\|\sum_{{\bf i}\in\mathcal{I}}[{\bf \Phi}]_{:, {\bf i}}[{\bf \Psi}_{k}]_{i_k, \hat{j}_k}\prod_{k'\in\mathcal{E}}[{\bf \Psi}_{k'}]_{i_{k'}, \hat{j}_{k'}}\prod_{k'\in\hat{\mathcal{E}}}[{\bf \Psi}_{k'}]_{i_{k'}, \hat{j}_{k'}}\|^2}\right).
\label{eq:ini}
\end{equation}
\begin{equation}
\label{eq:initialization_MP}
\max_{j_k}\left(\frac{\sum_{k''\in\hat{\mathcal{E}}}\sum_{i_{k''}=1}^{N_{k''}^{\rm s}}\|\sum_{k'''\in\mathcal{E}\cup\{k\}}\sum_{i_{k'''}=1}^{N_{k'''}^{\rm s}}[{\bf O}_{\rm \Phi}]_{:, {\bf i}}[{\bf \Psi}_{k}]_{i_k, \hat{j}_k}\prod_{k'\in\mathcal{E}}[{\bf \Psi}_{k'}]_{i_{k'}, \hat{j}_{k'}}\|^2}{\sum_{k''\in\hat{\mathcal{E}}}\sum_{i_{k''}=1}^{N_{k''}^{\rm s}}\|\sum_{k'''\in\mathcal{E}\cup\{k\}}\sum_{i_{k'''}=1}^{N_{k'''}^{\rm s}}[{\bf \Phi}]_{:, {\bf i}}[{\bf \Psi}_{k}]_{i_k, \hat{j}_k}\prod_{k'\in\mathcal{E}}[{\bf \Psi}_{k'}]_{i_{k'}, \hat{j}_{k'}}\|^2}\right).
\end{equation}
\setcounter{equation}{\value{MYtempeqncnt}}
\hrulefill
\vspace*{4pt}
\end{figure*}

\subsection{Multidimensional orthogonal matching pursuit (MOMP) algorithm}
Now we will create a multidimensional orthogonal matching pursuit algorithm to solve the multidimensional sparse recovery problem described in \eqref{eq:MP_multi}.
Every matching pursuit algorithm consists of two-stage iterations, namely  matching projection and residual update.
The classical orthogonal matching pursuit algorithm\cite{Tropp2007} starts by initializing the residual observation ${\bf O}_{\rm res}$ to ${\bf O}$. Then, at each iteration, it computes the dictionary value that maximizes the projection with the residual (matching projection). Next, the result of this step is used  to update the residual observation (residual update). In the next paragraphs we derive these two steps for the MOMP algorithm.

Using the classical formulation, the matching projection step has to solve $\max_{j}\left(\frac{\|{\bf O}_{\rm res}^{\rm H}[\overline{\bf \Phi}\overline{\bf \Psi}]_{:, j}\|^2}{\|[\overline{\bf \Phi}\overline{\bf \Psi}]_{:, j}\|^2}\right)$, which, with the multidimensional formulation, is equivalent to the expression
\begin{equation}\label{eq:maxproj}
\max_{{\bf j}\in\mathcal{J}}\left(\frac{\|{\bf O}_{\rm res}^{\rm H}\left(\sum_{{\bf i}\in\mathcal{I}}[{\bf \Phi}]_{:, {\bf i}}\prod_{k=1}^{N_{\rm D}}[{\bf \Psi}_k]_{i_k, j_k}\right)\|^2}{\|\sum_{{\bf i}\in\mathcal{I}}[{\bf \Phi}]_{:, {\bf i}}\prod_{k=1}^{N_{\rm D}}[{\bf \Psi}_k]_{i_k, j_k}\|^2}\right).
\end{equation}
By defining ${\bf O}_{\rm \Phi}\in\mathbb{C}^{N^{\rm m}\times N_{1}^{\rm s}\times\ldots\times N_{N_{\rm D}}^{\rm s}}$ as $[{\bf O}_{\rm \Phi}]_{:, {\bf i}} = {\bf O}_{\rm res}^{\rm H}[{\bf \Phi}]_{:, {\bf i}}$, the problem \eqref{eq:maxproj} can be rewritten as
\begin{equation}\label{eq:refinement_MP}
\max_{{\bf j}\in\mathcal{J}}\left(\frac{\|\sum_{{\bf i}\in\mathcal{I}}[{\bf O}_{\rm \Phi}]_{:, {\bf i}}\prod_{k=1}^{N_{\rm D}}[{\bf \Psi}_k]_{i_k, j_k}\|^2}{\|\sum_{{\bf i}\in\mathcal{I}}[{\bf \Phi}]_{:, {\bf i}}\prod_{k=1}^{N_{\rm D}}[{\bf \Psi}_k]_{i_k, j_k}\|^2}\right).
\end{equation}
This maximization step becomes a bottleneck in the iterative estimation process, since evaluating all possible combinations has a complexity of $O((N^{\rm q}+N^{\rm m})\prod_{k=1}^{N_{\rm D}}N_k^{\rm s}N_k^{\rm a})$.
Our proposed algorithm makes use of an iterative optimization by cuts instead, starting from a previous solution $(\hat{j}_1, \ldots, \hat{j}_{N_{\rm D}})\in\mathcal{J}$ that we want to refine for the index $j_{d}$ as
\begin{equation}
\max_{j_k}\left(\frac{\|\sum_{{\bf i}\in\mathcal{I}}[{\bf O}_{\rm \Phi}]_{:, {\bf i}}[{\bf \Psi}_{k}]_{i_k, \hat{j}_{d}}\prod_{k'=1, k'\neq k}^{N_{\rm D}}[{\bf \Psi}_{k'}]_{i_{k'}, \hat{j}_{k'}}\|^2}{\|\sum_{{\bf i}\in\mathcal{I}}[{\bf \Phi}]_{:, {\bf i}}[{\bf \Psi}_{k}]_{i_k, \hat{j}_k}\prod_{k'=1, k'\neq k}^{N_{\rm D}}[{\bf \Psi}_{k'}]_{i_{k'}, \hat{j}_{k'}}\|^2}\right).
\end{equation}
Evaluating this problem for the dictionary index $k$ only has  a computational complexity of $O((N^{\rm q}+N^{\rm m})N_k^{\rm a}\prod_{k'=1}^{N_{\rm D}}N_{k'}^{\rm s})$. Thus, operating with  every index-dependent dictionary independently has a complexity of $O\left((N^{\rm q}+N^{\rm m})\left(\sum_{k=1}^{N_{\rm D}}N_k^{\rm a}\right)\prod_{k=1}^{N_{\rm D}}N_k^{\rm s}\right)$, which is much lower than that of the original problem, i.e. $O((N^{\rm q}+N^{\rm m})\prod_{k=1}^{N_{\rm D}}N_k^{\rm s}N_k^{\rm a})$. 
Note that this technique requires an initial estimation that is later refined in multiple iterations. Therefore, an additional initialization step is required.

Our solution for the initialization for the dictionary index $k$ starts by considering the set $\mathcal{E}$ of dictionary indexes $k'$ for which we have an estimation $\hat{j}_{k'}$.
Note that, since this is meant for initialization purposes only, $k\notin\mathcal{E}$.
Analogously, we define $\hat{\mathcal{E}}$ as the set of indexes $k'$ for which we do not have an estimation $\hat{j}_{k'}$ excluding $k$.
We can now split the indexes among different sets, leaving the formulation as shown in \eqref{eq:ini}.
Then, because we do not have an estimation of the values $\hat{j}_{k'}$, we move to approximate the quantity in \eqref{eq:ini}. We define ${\bf \Gamma}\in\mathbb{C}^{\bigotimes_{k=1}^{N_{\rm D}}N_k^{\rm s}}$ as $[{\bf \Gamma}]_{\bf j} = \prod_{k'\in\hat{\mathcal{E}}}[{\bf \Psi}_{k'}]_{i_{k'}, \hat{j}_{k'}}$. We relax the constraint on $k'$, and then apply the inequality $\frac{\sum_ia_i}{\sum_ib_i}\leq\sum_i\frac{a_i}{b_i}$ for positive numbers $\{a_i\}$ and $\{b_i\}$ over the expansions of the Frobenius norms corresponding to the terms that contain ${\bf \Gamma}$, to get the relaxation of \eqref{eq:ini} to the expression in \eqref{eq:initialization_MP}.
Although this problem is only a relaxation, note that we are only trying to find an initialization for the iterative process, so it will still be possible to find later the solution to the original multidimensional orthogonal matching pursuit problem.
Due to the same reason, for implementation purposes, it is possible to replace the dictionaries with other dictionaries with less atoms, and even ignore computing the coefficient if we expect not to have a high variation of values in the denominator.

\begin{figure*}[t!]
\begin{center}
\includegraphics[width=0.85\linewidth]{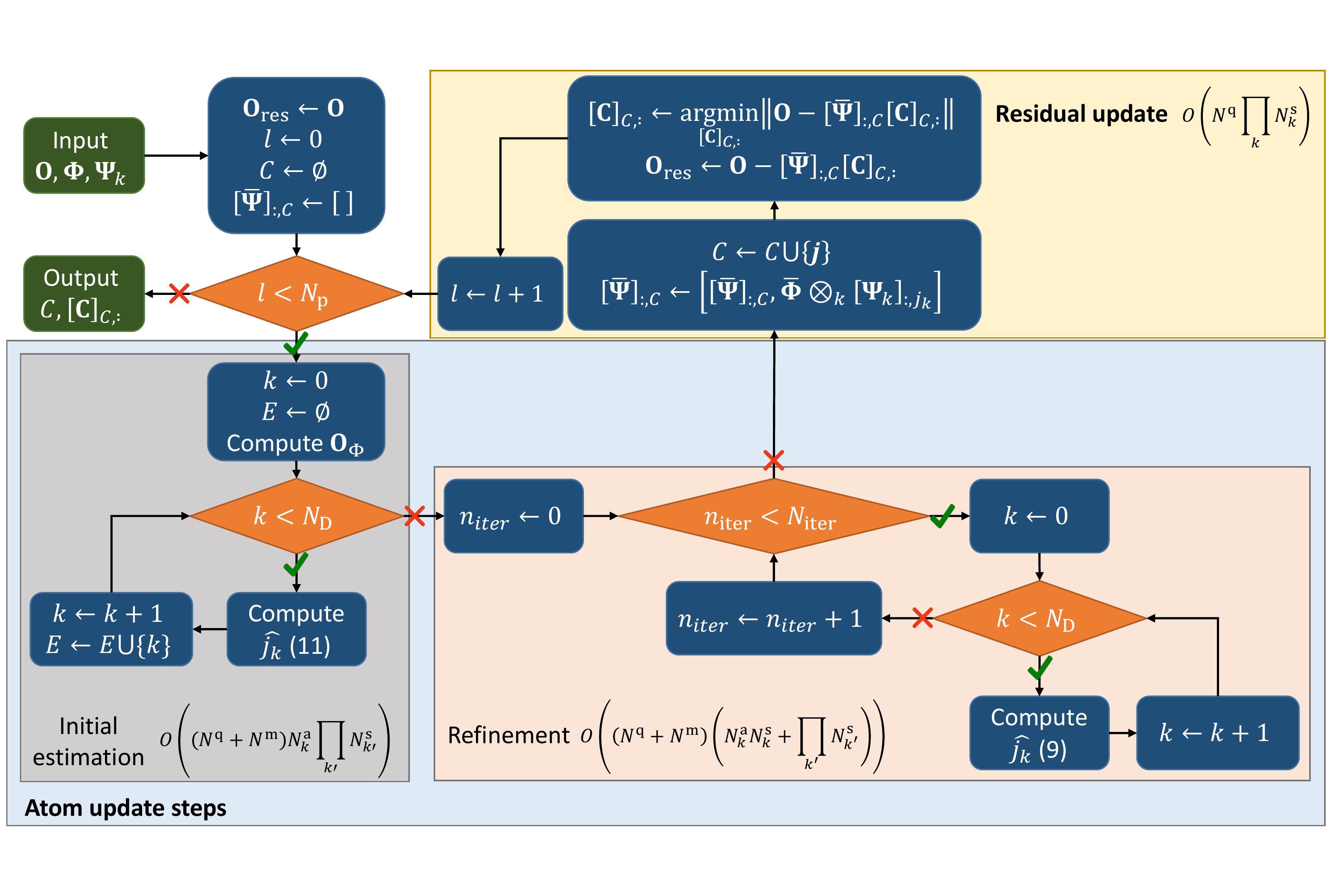}
\end{center}
\caption{Block diagram for the MOMP algorithm.}
\label{fig:alg_diag}
\end{figure*}

Next, we need to adapt the classical residual update strategy to operate with multiple dictionaries.
In the conventional version of OMP, after the matching projection step, we would include the dictionary index $\hat{j}$ in the support set $\mathcal{C}$, and compute ${\bf O}_{\rm res}$ as ${\bf O}-[\overline{\bf \Phi\Psi}]_{:, \overline{\mathcal{C}}}[\overline{\bf C}]_{\overline{\mathcal{C}}, :}$ being $[\overline{\bf C}]_{\overline{\mathcal{C}}, :}$ the solution to the linear minimum mean square error problem $\min_{[\overline{\bf C}]_{\overline{\mathcal{C}}, :}}\|{\bf O}-[\overline{\bf \Phi}\overline{\bf \Psi}]_{:, \overline{\mathcal{C}}}[\overline{\bf C}]_{\overline{\mathcal{C}}, :}\|^2$.
Our solution consists on building the equivalent $[\overline{\bf \Phi\Psi}]_{:, \overline{\mathcal{C}}}$ matrix for the classical problem by computing its last column at each step and appending it.
Considering $\hat{\bf j}$ as the index of the last estimated component added to $\mathcal{C}$, the last column of $[\overline{\bf \Phi\Psi}]_{:, \overline{\mathcal{C}}}$ is given by $\sum_{{\bf i}\in\mathcal{I}}{\bf \Psi}_{:, {\bf i}}\prod_{k=1}^{N_{\rm D}}[{\bf \Psi}_{k}]_{i_k, \hat{j}_k}$.
Knowing that $[\overline{\bf C}]_{\overline{\mathcal{C}}, :}=[{\bf C}]_{\mathcal{C}, :}$ we can solve the residual update.
Due to the sparse nature of ${\bf C}$ it is simpler  to report $\mathcal{C}$ and $[\overline{\bf C}]_{\mathcal{C}, :}$ rather than ${\bf C}$. The multi-dimensional orthogonal matching pursuit algorithm is now complete.
Its final form is included as a block diagram in Fig.~\ref{fig:alg_diag}, together with the complexities associated to each step, while Algorithm 1 contains the pseudocode.  A python implementation of MOMP can be downloaded from \cite{CodeMOMPCore}.
\begin{algorithm}
    \caption{Multidimensional Matching Pursuit (MOMP) Algorithm}
    \label{alg:MP-cuts}
\begin{algorithmic}
\STATE ${\bf O}_{\rm res}\gets {\bf O}$
\STATE Initialize $\mathcal{C}\gets\emptyset$ and $[\overline{\bf \Psi}]_{:, \mathcal{C}}\gets []$
\FOR{$s\leq N^{\rm sp}$}
 \STATE Compute the ${\bf O}_{\rm \Phi}$ required for equation \eqref{eq:refinement_MP}
 \STATE Initialize $\mathcal{E}\gets \emptyset$
 \FOR{$d\leq N_{\rm D}$}
  \STATE Compute an initial estimation of $\hat{j}_{d}$ using equation \eqref{eq:initialization_MP}
  \STATE $\mathcal{E}\gets\mathcal{E}\cup\{d\}$
 \ENDFOR
 \FOR{$n_{\rm iter} \leq N_{\rm iter}$}
  \FOR{$d\leq N_{\rm D}$}
   \STATE Refine the estimation of $\hat{j}_{d}$ using equation \eqref{eq:refinement_MP}
  \ENDFOR
 \ENDFOR
 \STATE Update $\mathcal{C}\gets \mathcal{C}\bigcup\{[\hat{j}_1, \ldots, \hat{j}_{N_{\rm D}}]\}$
 \STATE Update $[\overline{\bf \Psi}]_{:, \mathcal{C}}\gets [[\overline{\bf \Psi}]_{:, \mathcal{C}}, \bigotimes_{d = 1}^{N_{\rm D}}[{\bf \Psi}_{k}]_{:, \hat{j}_d}]$
 \STATE Compute $[\overline{\bf C}]_{\mathcal{C}, :}$ achieving $\min\|{\bf O}-[\overline{\bf \Psi}]_{:, \mathcal{C}}[\overline{\bf C}]_{\mathcal{C}, :}\|^2$
 \STATE Update ${\bf O}_{\rm res}\gets {\bf O}-[\overline{\bf \Psi}]_{:, \mathcal{C}}[\overline{\bf C}]_{\mathcal{C}, :}$
\ENDFOR
\STATE Output $\mathcal{C}, [\overline{\bf C}]_{\mathcal{C}, :}$
\end{algorithmic}
\end{algorithm}

\section{MOMP-based mmWave channel estimation}\label{sec:channel_estimation}
Different variations of the matching pursuit strategy, including OMP and SOMP \cite{Tropp2007,Tropp2005}, have been applied in recent work to estimate the channel at millimeter wave bands. In this section, we review first the compressive formulation of the mmWave channel estimation problem adopted in previous work, and identify the limitations associated to the proposed construction for the dictionary, which leads to a high computational complexity in the sparse recovery process. Then, we derive a new formulation based on MOMP which leverages independent dictionaries in the angular and delay domains without constructing a global dictionary of larger size.  

\subsection{Background}
Communication at mmWave frequencies requires the usage of large antenna arrays both at the transmitter and the receiver. These arrays have to be configured as a function of the channel, which is generally unknown and needs to be estimated frequently.
When we consider a frequency selective channel, it is defined by a sequence of matrices $\{{\bf H}_d\}$, $d=1,\ldots D$, being $d$ the index that loops over the delay taps, and $D$ the delay tap length \cite{Venugopal2017}.
${\bf H}_d$ 
can be written in terms of the channel path properties and the geometry of the transmit and receive antenna arrays using a geometric channel model with $L$ paths \cite{Venugopal2017}.
Each path $l$ has a complex gain $\alpha_l$, including the signal phase and the path loss, a direction of arrival ${\boldsymbol \theta}_l$, a direction of departure ${\boldsymbol \phi}_l$, and a delay $\tau_l$. The angular directions can be represented using polar coordinates, but to simplify notations and derivations, we will use vector directions, i.e. ${\boldsymbol \theta}_l, {\boldsymbol \phi}_l \in \{{\bf v}\in\mathbb{R}^3\text{ such that }\|{\bf v}\|=1\}$.
The time that passes between the beginning of the transmission and the beginning of the reception is $\tau_0$, and the sampling time is denoted as $T_{\rm s}$.
The physical properties of the devices and the transmitted signal in the direction of arrival ${\boldsymbol \theta}$ and direction of departure ${\boldsymbol \phi}$ at time $t$ are encompassed, respectively, in the steering vectors ${\bf a}_{\rm R}({\boldsymbol \theta})$ and ${\bf a}_{\rm T}({\boldsymbol \phi})$,  and the bandlimited pulse shaping function $p(t)$, that includes  filtering effects at the transmitter and receiver.
The sizes of the steering vectors are equal to the number of antenna elements in the transmit and receive array, $\NTX$ and $\NRX$, respectively, when considering linear arrays.  
With these definitions, the expression for the $d$-th delay tap of the channel is
\setcounter{equation}{11}
\begin{equation}\label{eq:geometric_channel_no_time_response}
{\bf H}_{d} = \sum_{l = 1}^{L}\alpha_l{\bf a}_{\rm R}({\boldsymbol \theta}_l){\bf a}_{\rm T}^{\rm H}({\boldsymbol \phi}_l)p((d-1)T_{\rm s}+\tau_0-\tau_l).
\end{equation}
Similarly to the steering vector, we can define a time response vector ${\bf a}_{\rm D}(\tau-\tau_0)\in\mathbb{C}^{D}$ as $[{\bf a}_{\rm D}(\tau-\tau_0)]_d = p((d-1)T_{\rm s}+\tau_0-\tau)$ , so that the expression for the channel can be written in a more compact way as
\begin{equation}\label{eq:geometric_channel}
{\bf H}_{d} = \sum_{l = 1}^{L}\alpha_l{\bf a}_{\rm R}({\boldsymbol \theta}_l){\bf a}_{\rm T}^{\rm H}({\boldsymbol \phi}_l)[{\bf a}_{\rm D}(\tau_l-\tau_0)]_d.
\end{equation}

The channel matrices ${\bf H}_{d}$ can be modeled as sparse at mmWave wave frequencies \cite{mmwavechannel}, which paves the way for the connection between channel estimation strategies and sparse recovery algorithms. Extensive prior work has exploited this sparse structure to define effective mmWave channel estimation algorithms, either in the time domain or the frequency domain \cite{Venugopal2017,SWOMP2017,lee2014exploiting,Coma2018, Wu2019, Zhu2019}, by solving a given sparse recovery problem. In this context, the measurements correspond to different instances of the receive signal when transmitting different training symbols with different combinations of training precoders and combiners; the measurement matrix depends on the sequence of training beamformers used at the transmitter and the receiver during the training process; and finally, the dictionary usually employed depends on the pulse shaping filter if considered, and the array steering vectors of the transmitter and receiver evaluated on a grid of possible directions of arrival and departure.  For example, in the formulation proposed in \cite{Venugopal2017}, a hybrid mmWave architecture with $\NTX$ transmit antennas, $\NRX$ receive antennas, and $M_{\text{RF}}$ chains at both sides is considered. The training symbol vector at instance $n$ for the $m$-th training frame is defined as $\bs_m[n]$, the corresponding training precoder is the matrix  ${\bF}_m$, and the corresponding training combiner is ${\bW}_m$. The frame length is denoted as $N$. With these definitions, the received signal for the $m$-th training frame can be written as
\begin{equation} 
{\by}_m =\left(\SmBlc \left(\bI_{D}\otimes \bF^{\rm T}_m\right) \otimes {\bW}_m^{\rm H}\right)\hc + {\bee}_{m},
\end{equation}
being $\bee_m$ the postcombining noise, $\bS_m \in \mathbb{C}^{N\times D M_\text{RF}}$ the matrix built from the $m$-th frame training symbols as
\begin{align}
&~\SmBlc = \begin{bmatrix}
\bs^{\rm T}_m[1] & 0  & \cdots &0 \\ 
\bs^{\rm T}_m[2] & \bs^{\rm T}_m[1] & \cdots & .\\ 
\vdots & \vdots & \ddots &\vdots \\ 
\bs^{\rm T}_m[N] & \cdots & \cdots & \bs^{\rm T}_m[{N-D+1}].
\end{bmatrix}, \label{eqn:block_Sm}
\end{align}
and $\hc$ the vectorized channel defined as
\begin{align}
\hc = \begin{bmatrix}
\vect(\bH_1)\\
\vect(\bH_2)\\
\vdots \\
\vect(\bH_{D})
\end{bmatrix}. \label{eqn:vec_channel}
\end{align} 
A sparse vectorized version of the channel ${\bx}_{\td}$ such that $\hc={\mathbf{\Psi}_{\td}\bx}_{\td}$, where $\mathbf{\Psi}_{\td}$ is the sparsifying dictionary, is then estimated by solving with OMP the sparse recovery problem
\begin{equation}
 \underset{\bx_{\td}}{\min} \Vert \by_{\td}  - {\bf{\Phi}}_{\td} {\bf{\Psi}}_{\td}\bx_{\td} \Vert_2, \label{eqn:cs_fund_cvx}
\end{equation} 
where the mesurement vector $\by_{\td} \in {\bbC}^{NM_{\rm TR}M_\text{RF}^2 \times 1}$ contains the $M_{\rm TR}$ received training frames $\by_{m_{\rm TR}}$ stacked as
\begin{equation}
\by_{\td} = \begin{bmatrix}
\by_1\\
\by_2\\
\vdots \\
\by_{M_{\rm TR}}
\end{bmatrix}.
\end{equation}
The measurement matrix $\mathbf{\Phi}_{\td}  \in {\bbC}^{NM_{\rm TR}M_\text{RF}^2\times D\NRX\NTX}$ can be written as \cite{Venugopal2017}
\begin{align}
\mathbf{\Phi}_{\td} = \begin{bmatrix}
\SBlc \left(\bI_{D}\otimes \bF^{\rm T}_1\right)\otimes {\bW}_1^{\rm H}\\
{\mathbf{S}}_{2}\left(\bI_{D}\otimes \bF^{\rm T}_2\right)\otimes {\bW}_2^{\rm H}\\
\hspace{-0.07in}\vdots\\
{\mathbf S}_{M_{\rm TR}} \left(\bI_{D}\otimes \bF^{\rm T}_{M_{\rm TR}}\right)\otimes {\bW}_{M_{\rm TR}}^{\rm H}
\end{bmatrix}. \label{eqn:sensemat}
\end{align}
Regarding the dictionary ${\bf{\Psi}}_{\td}$, it is defined from the matrices $\Atx \in \NTX \times \Gt$, $\Arx \in \NRX \times \Gr$ and $\bp_d$, where $\Gt$ and $\Gr$ are the grid sizes for the AoD and AoA.  $\Atx$ consists of columns $\at(\tilde{\phi})$, with $\tilde{\phi}$ drawn from the grid for the AoD , while $\Arx$ consists of columns $\ar(\tilde{\theta})$, with $\tilde{\theta}$ drawn from the grid for the AoA. Vectors $\bp_d$ are defined as the sampled version of the pulse shaping filter, having entries ${p}_d(n) = p\left( (d-n\frac{D}{\Gc})\Ts\right) $, for $d=1,2,~\cdots,~D$ and $n=1,2,~\cdots,~\Gc$.
Note that $\Atx$ and $\Arx$ are dictionaries for the angle of departure and arrival respectively, while  $\bp_d$ will allow the representation of different delays from a quantized grid of size $\Gc$.  
The expression of the dictionary matrix is 
\begin{equation}
\mathbf{\Psi}_{\td}= \begin{bmatrix}
\left( {\Atxbar}\otimes \Arx \right)\otimes{\bp}^{\rm T}_1\\
\left( {\Atxbar}\otimes \Arx \right)\otimes{\bp}^{\rm T}_2\\
\hspace{-0.07in}\vdots\\
\left( {\Atxbar}\otimes \Arx \right)\otimes{\bp}^{\rm T}_{D}
\end{bmatrix} \in {\bbC}^{D\NRX\NTX \times \Gc\Gr\Gt}. \label{eqn:td_dictionary}
\end{equation}
Note that when solving the sparse recovery problem in \eqref{eqn:cs_fund_cvx}, the location of the non zero coefficients will point to a given column in the dictionary matrix, so that the angle of departure, angle of arrival and delay for each path can be identified. 

{\bf Remark:} {\em The main limitation of the sparse estimation algorithm described above comes from the structure of the dictionary matrix. On the one hand, this dictionary matrix is obtained as the Kronecker product of dictionaries for the angle of departure, angle of arrival, and delay, so the size of the global dictionary depends on the product of the sizes of these dictionaries. On the other hand, the accuracy on the reconstruction of the mmWave channel heavily depends on the grid selected to build the dictionary. Finer grids and larger arrays lead to larger dictionaries and higher accuracies, what leads to an increased complexity of the reconstruction process when combined with the structure of the dictionary marix.}

Most of the algorithms for channel estimation at mmWave are based in this type of dictionary matrix. Thus, one limitation of existing algorithms is their high complexity, which comes from the large size of the dictionary used to obtain a good performance. This limitation is even more critical when an additional estimate of the position and orientation of the user is desired. In this case, under some mild assumptions, the estimated channel parameters can be transformed into the localization information, but the sensitivity of this step to channel estimation errors is high, and higher accuracies (i.e. larger dictionaries) are needed than when the channel estimate is obtained with the only goal of designing the beamformers to maximize the communication performance. With this motivation in mind, in the next section we formulate the mmWave channel estimation problem in the time domain as a sparse recovery problem with a multidimensional dictionary, that can be solved with the MOMP algorithm defined in Section~\ref{sec:MOMP}. This formulation allows a higher resolution in the dictionary without having to resort to increasing the dictionary size, since it is not  
obtained as a Kronecker product of dictionaries for every sparse domain. 

\subsection{MOMP-based formulation}
We consider an up-link scenario in which the users operate either with an analog beamforming MIMO architecture, or a hybrid architecture. Note that the analog beamforming architecture is a special case of the hybrid one when operating with a single RF chain. In both cases, the users employ a $N_{\rm T}^{\rm x}\times N_{\rm T}^{\rm y}$ uniform rectangular antenna array, while the access point makes use of a  $N_{\rm R}^{\rm x}\times N_{\rm R}^{\rm y}$ uniform rectangular array antenna. The number of RF chains at the access point is denoted as $M_{\rm R}$, while the number of RF chains for the user is denoted as $M_{\rm T}$.
The observation consists of $M$ received training frames, with $Q$ vector symbols of size $N_{\rm S}$ per frame, each frame obtained using a different beampattern pair.
For each measurement $m$, the $Q$ training vector symbols are zero padded, so that the pilot signal can be defined as  ${\bf S}_m\in\mathbb{C}^{N_{\rm S}(Q+D)}$, where $D$  denotes the number of zeros being added.
During training, we choose $N_{\rm S}=M_{\rm T}$, what leads to a square digital precoder.
Regarding the digital combiner, it is also square, so that the information acquired by the receive RF chains is not further compressed. 
Thus, the digital training precoders/combiners for the $m$-th training frame can be written as ${\bf F}_m^{\rm BB} \in\mathbb{C}^{M_{\rm T}\times M_{\rm T}}$ and ${\bf W}_m^{\rm BB}\in\mathbb{C}^{M_{\rm R}\times M_{\rm R}}$.
For the analog precoding/combining stage for the $m$-th training frame we use the filters ${\bf F}_m^{\rm RF} \in\mathcal{U}^{N_{\rm T} \times M_{\rm T}}$  and ${\bf W}_m^{\rm RF} \in \mathcal{U}^{N_{\rm R}\times M_{\rm R}}$, respectively. This way, the hybrid precoders/combiners are  ${\bf F}_m={\bf F}_m^{\rm RF}{\bf F}_m^{\rm BB}\in\mathbb{C}^{N_{\rm T}\times M_{\rm T}}$ and ${\bf W}_m={\bf W}_m^{\rm RF}{\bf W}_m^{\rm BB}\in\mathbb{C}^{N_{\rm R}\times M_{\rm R}}$.
Finally, we define the transmit power $P_{\rm t}$ and the noise matrix ${\bf N}_m\in\mathbb{C}^{M_{\rm R}\times Q}$ of independent identically distributed Gaussian entries with distribution $\mathcal{N}(0, \sigma^2)$.

With these definitions, the received signal during training, ${\bf Y}_m\in\mathbb{C}^{M_{\rm R}\times Q}$ is 
\begin{equation}\label{eq:general_channel_meas_w_channel_t}
[{\bf Y}_m]_{:, q} = \sqrt{P_{\rm t}}\sum_{d = 1}^D{\bf W}_m^{\rm H}{\bf H}_d{\bf F}_m[{\bf S}_m]_{:, q+D-d} + {\bf W}_m^{\rm H}[{\bf N}_m]_{:, q}.
\end{equation}
The whitened version of \eqref{eq:general_channel_meas_w_channel_t} can be obtained from the Cholesky decomposition of the noise covariance  matrix $\bR_n$ as in \cite{SWOMP2018}, i.e. $\bR_n={\bf L}_m{\bf L}_m^{\rm H}={\bf W}_m^{\rm H}{\bf W}_m$, as
\begin{equation}\label{eq:general_channel_meas_w_channel_white}
{\bf Y}'_m={\bf L}_m^{-1}{\bf Y}_m. 
\end{equation}
We define now ${\bf W}'_m = {\bf W}_m({\bf L}_m^{-1})^{\rm H}$ and ${\bf N}'_m = {\bf L}_m^{-1}{\bf W}_m^{\rm H}{\bf N}_m$. Note that by construction,  ${\bf N}'_m$ also contains white noise.
If we substitute \eqref{eq:geometric_channel} in \eqref{eq:general_channel_meas_w_channel_white}, we obtain
\begin{equation}\label{eq:general_channel_meas}
\begin{split}
{\bf Y}'_m = \sqrt{P_{\rm t}}\sum_{d = 1}^{D}({\bf W}'_m)^{\rm H}{\bf H}_d{\bf F}_m[{\bf S}_m]_{:, q+D-d}[{\bf a}_{\rm D}^{\rm T}(\tau_l-\tau_0)]_d \\ + {\bf N}'_m.
\end{split}
\end{equation}

To define the expressions for the steering vectors, we  focus on the $N_{\rm R}^{\rm x}\times N_{\rm R}^{\rm y}$ planar array at the receiver first. Then, we define  the $(n_{\rm R}^{\rm x}N_{\rm R}^{\rm y}+n_{\rm R}^{\rm y})$-th antenna element located in position $\frac{\lambda}{2}[n_{\rm R}^{\rm x}-1, n_{\rm R}^{\rm y}-1, 0]$.Assuming ${\boldsymbol \theta}$ and ${\boldsymbol \phi}$ to be the unitary vectors $(\theta_{\rm x}, \theta_{\rm z}, \theta_{\rm z})$ and $(\phi_{\rm x}, \phi_{\rm z}, \phi_{\rm z})$, the receive steering vector can be written as 
$[{\bf a}_{\rm R}({\boldsymbol \theta})]_{n_{\rm R}^{\rm x}N_{\rm R}^{\rm y}+n_{\rm R}^{\rm y}} = e^{-j\pi((n_{\rm R}^{\rm x}-1)\phi_{\rm x}+(n_{\rm R}^{\rm y}-1)\phi_{\rm y})}$.  The same type of description can be done for the planar array at the transmit side, simply by substituting the subindex R by T and $\theta$ by $\phi$. 
These expressions for the array steering vectors  can be decomposed as
\begin{equation}\label{eq:steering_decomposition}
\begin{array}{rl}
{\bf a}_{\rm R}({\boldsymbol \theta}) = & {\bf a}_{\rm R}^{\rm x}(\theta_{\rm x})\otimes{\bf a}_{\rm R}^{\rm y}(\theta_{\rm y}),\\
{\bf a}_{\rm T}({\boldsymbol \phi}) = & {\bf a}_{\rm T}^{\rm x}(\phi_{\rm x})\otimes{\bf a}_{\rm T}^{\rm y}(\phi_{\rm y}),
\end{array}
\end{equation}
being the partial steering vectors ${\bf a}_{\rm R}^{\rm x}(\theta_{\rm x})\in\mathbb{C}^{N_{\rm R}^{\rm x}}$, ${\bf a}_{\rm R}^{\rm y}(\theta_{\rm y})\in\mathbb{C}^{N_{\rm R}^{\rm y}}$, ${\bf a}_{\rm T}^{\rm x}(\phi_{\rm x})\in\mathbb{C}^{N_{\rm T}^{\rm x}}$ and ${\bf a}_{\rm T}^{\rm y}(\phi_{\rm y})\in\mathbb{C}^{N_{\rm T}^{\rm y}}$ with expressions $[{\bf a}_{\rm R}^{\rm x}(\theta_{\rm x})]_{n_{\rm R}^{\rm x}} = e^{-j\pi(n_{\rm R}^{\rm x}-1)\theta_{\rm x}}$, $[{\bf a}_{\rm R}^{\rm y}(\theta_{\rm y})]_{n_{\rm R}^{\rm y}} = e^{-j\pi(n_{\rm R}^{\rm y}-1)\theta_{\rm y}}$, $[{\bf a}_{\rm T}^{\rm x}(\phi_{\rm x})]_{n_{\rm T}^{\rm x}} = e^{-j\pi(n_{\rm T}^{\rm x}-1)\phi_{\rm x}}$ and $[{\bf a}_{\rm T}^{\rm y}(\phi_{\rm y})]_{n_{\rm T}^{\rm y}} = e^{-j\pi(n_{\rm T}^{\rm y}-1)\phi_{\rm y}}$.

\begin{table*}[t!]
\centering
\begin{tabular}{|c|l|}
\hline
Method & Complexity order \\
\hline
OMP & $\mathcal{O}((M_{\rm R}^{\rm x}M_{\rm R}^{\rm y}M_{\rm T}^{\rm x}M_{\rm T}^{\rm y}D)^2K_{\rm res}^5)$\\
\hline
MOMP & $\mathcal{O}((M_{\rm R}M_{\rm T}D)^2+(1+M_{\rm R}M_{\rm T}D)((M_{\rm R}^{\rm x})^2+(M_{\rm R}^{\rm y})^2+(M_{\rm T}^{\rm x})^2+(M_{\rm T}^{\rm y})^2+D^2)K_{\rm res})$\\
\hline
\end{tabular}\caption{Complexity orders of OMP and MOMP when applied to the mmWave channel estimation problem.}\label{tab:complexity}
\end{table*}

Since our objective is to reconstruct the channel, sparsity appears in the angular and delay dimensions, and the x- and y-dimensions of the angular domains are independent, our obvious choice for the dictionaries is \cite{Palacios2022Eusipco}.
\begin{align}\label{eq:exp_X}
{\bf \Psi}_{1} & = [{\bf a}_{\rm R}^{\rm x}(\overline\theta_1^{\rm x}), \ldots, {\bf a}_{\rm R}(\overline{\theta}_{N_{1}^{\rm a}}^{\rm x})],\\
{\bf \Psi}_{2} & = [{\bf a}_{\rm R}^{\rm y}(\overline\theta_1^{\rm y}), \ldots, {\bf a}_{\rm R}(\overline{\theta}_{N_{2}^{\rm a}}^{\rm y})],\\
{\bf \Psi}_{3} & = [{\bf a}_{\rm T}^{\rm x}(\overline\phi_1^{\rm x})^{\rm H}, \ldots, {\bf a}_{\rm T}(\overline{\phi}_{N_{3}^{\rm a}}^{\rm x})^{\rm H}],\\
{\bf \Psi}_{4} & = [{\bf a}_{\rm T}^{\rm y}(\overline\phi_1^{\rm y})^{\rm H}, \ldots, {\bf a}_{\rm T}(\overline{\phi}_{N_{4}^{\rm a}}^{\rm y})^{\rm H}],\\
{\bf \Psi}_{5} & = [{\bf a}_{\rm D}(\overline{\tau}_1), \ldots, {\bf a}_{\rm D}(\overline{\tau}_{N_{5}^{\rm a}})],
\end{align}
when considering $\{\overline\theta_1^{\rm x}, \ldots, \overline{\theta}_{N_{1}^{\rm a}}^{\rm x}\}$, $\{\overline\theta_1^{\rm y}, \ldots, \overline{\theta}_{N_{2}^{\rm a}}^{\rm y}\}$, $\{\overline\phi_1^{\rm x}, \ldots, \overline{\theta}_{N_{3}^{\rm a}}^{\rm x}\}$, $\{\overline{\phi}_1^{\rm y}, \ldots, \overline{\theta}_{N_{4}^{\rm a}}^{\rm y}\}$ and $\{\overline{t}_1, \ldots, \overline{t}_{N_{5}^{\rm a}}\}$ to be discrete versions of the domains for $\theta_{\rm x}$, $\theta_{\rm y}$, $\phi_{\rm x}$, $\phi_{\rm y}$ and $\tau-\tau_0$ variables respectively.
This implies $N_{\rm D}=5$, $N_{1}^{\rm s}=N_{\rm R}^{\rm x}$, $N_{2}^{\rm s}=N_{\rm R}^{\rm y}$, $N_{3}^{\rm s}=N_{\rm T}^{\rm x}$, $N_{5}^{\rm s}=N_{\rm T}^{\rm y}$, $N_{5}^{\rm s}=D$.
With this multidimensional dictionary configuration, and ignoring quantization effects caused by the finite resolution of the dictionaries, we can define ${\bf C}$ as
\begin{equation}\label{eq:exp_S}
[{\bf C}]_{{\bf j}}=\left\lbrace\begin{array}{cl}
\alpha_l & \text{if } \begin{array}{ccc}
\theta_l^{\rm x} & = & \overline{\theta}_{j_1}^{\rm x}\\
\theta_l^{\rm y} & = & \overline{\theta}_{j_2}^{\rm y}\\
\phi_l^{\rm x} & = & \overline{\phi}_{j_3}^{\rm x}\\
\phi_l^{\rm y} & = & \overline{\phi}_{j_4}^{\rm y}\\
\tau_l & = & \overline\tau_{j_5}
\end{array}\\
0 & \text{otherwise}
\end{array}\right.
\end{equation}
The channel expression can be finally written as
\begin{multline}\label{eq:exp_H_chest}
[{\bf H}_d]_{n_{\rm R}^{\rm x}N_{\rm R}^{\rm y}+n_{\rm R}^{\rm y}, n_{\rm T}^{\rm x}N_{\rm T}^{\rm y}+n_{\rm T}^{\rm y}} =\\
 \sum_{{\bf j}\in\mathcal{J}}[{\bf \Psi}_1]_{n_{\rm R}^{\rm x}, j_1}[{\bf \Psi}_2]_{n_{\rm R}^{\rm y}, j_2}[{\bf \Psi}_3]_{n_{\rm T}^{\rm x}, j_3}[{\bf \Psi}_4]_{n_{\rm T}^{\rm y}, j_4}[{\bf \Psi}_5]_{d, j_k}{\bf C}{{\bf j}},
\end{multline}
for $n_{\rm R}^{\rm x}\leq N_{\rm R}^{\rm x}, n_{\rm R}^{\rm y}\leq n_{\rm R}^{\rm y}, n_{\rm T}^{\rm x}\leq N_{\rm T}^{\rm x}, n_{\rm T}^{\rm y}\leq n_{\rm T}^{\rm y}$ and $d\leq D$.
Since our purpose is to transform the channel estimation problem into the multi-dimensional sparse estimation problem, the next step is to relabel the sub-index by their dictionary counterparts, that is substituting $(n_{\rm R}^{\rm x},n_{\rm R}^{\rm y}, n_{\rm T}^{\rm x},n_{\rm T}^{\rm y}, d)$ with ${\bf i} = (i_1, i_2, i_3, i_4, i_5)$ to get the cleaner formulation
\begin{equation}\label{eq:exp_H}
[{\bf H}_{i_5}]_{i_1N_{\rm R}^{\rm y}+i_2, i_3N_{\rm T}^{\rm y}+i_4} = \sum_{{\bf j}\in\mathcal{J}}\prod_{k=1}^{N_{\rm D}}[{\bf \Psi}_k]_{i_k, j_k}{\bf C}{{\bf j}},
\end{equation}
for every combination of $i_k\leq N_k^{\rm s}\quad\forall k\leq N_{\rm D}$.
This allows us to rewrite eq.~\eqref{eq:general_channel_meas} as
\begin{multline}\label{eq:general_channel_meas_partialMOMP}
{\bf Y}'_m = \sqrt{P_{\rm t}}\sum_{{\bf i}\in\mathcal{I}}
({\bf W}'_m]_{i_1N_{\rm R}^{\rm y}+i_2, :}^{\rm H}[{\bf F}_m]_{i_3N_{\rm T}^{\rm y}+i_4, :}\\
[{\bf S}_m]_{:, q+D-i_5}[{\bf a}_{\rm D}^{\rm T}(\tau_l-\tau_0)]_{i_5}\sum_{{\bf j}\in\mathcal{J}}(\prod_{k=1}^{N_{\rm D}}[{\bf \Psi}_k]_{i_k, j_k}{\bf C}{{\bf j}})) \\
+ {\bf N}'_m.
\end{multline}
Note that every measurement entry is independent from the other ones, since no training configuration is repeated.
Therefore the observation matrix will have size ${\bf O}\in\mathbb{C}^{MM_{\rm R}Q\times 1}$.
In consequence, the first index of ${\bf O}$ must run over all measurement entries, i.e.  $[{\bf O}]_{mM_{\rm R}Q+m_{\rm R}Q+q, 1}=[{\bf Y}'_m]_{m_{\rm R}, q}$.
From \eqref{eq:general_channel_meas_partialMOMP} and this definition of ${\bf O}$, it follows that the weight of the contribution of $\sum_{{\bf j}\in\mathcal{J}}\prod_{k=1}^{N_{\rm D}}[{\bf \Psi}_k]_{i_k, j_k}{\bf C}{{\bf j}}$ to $[{\bf O}]_{mM_{\rm R}Q+m_{\rm R}Q+q, 1}=[{\bf Y}'_m]_{m_{\rm R}, q}$ is
 \begin{multline}\label{eq:exp_A}
[{\bf \Phi}\hspace*{-0.5mm}]_{mM_{\rm R}Q+m_{\rm R}Q+q, {\bf i}} \hspace*{-0.5mm}=\\  \hspace*{-1mm} [{\bf W}'_m]_{i_1N_{\rm R}^{\rm y}+i_2, m_{\rm R}}^*\hspace*{-1mm}[{\bf F}_m{\bf S}_m]_{i_3N_{\rm R}^{\rm y}+i_4, q+D-i_5}.
\end{multline}
Taking into account that the noise is white, the maximum likelihood estimator is given by the minimum mean square estimator which is the solution to eq.~\eqref{eq:MP_multi}, and we can solve it using the MOMP algorithm for a sparse solution.

Once we apply the multi-dimensional matching pursuit algorithm to this problem, we obtain the parameters of $N^{\rm sp}$ paths in the form of $\mathcal{C}$ and $[\overline{\bf C}]_{\mathcal{C}, :}$.
We start by subtracting $\alpha_l{\bf a}_{\rm T}({\boldsymbol \phi}_l)^{\rm H}{\bf f}_{m_{\rm T}} = [[\overline{\bf C}]_{\mathcal{C}, :}]_{l, m_{\rm T}}$.
Then, for $(j_1, j_2, j_3, j_4, j_5)$ being the $l$-th element of $\mathcal{C}$, we can obtain $\theta_l^{\rm x} = \overline{\theta}_{j_1}^{\rm x}$, $\theta_l^{\rm y} = \overline{\theta}_{j_2}^{\rm y}$, $\phi_l^{\rm x} = \overline{\phi}_{j_3}^{\rm x}$, $\phi_l^{\rm y} = \overline{\phi}_{j_4}^{\rm y}$ and $\tau_{l}-\tau_0 = \overline{t}_{j_5}$. Then we compute ${\theta}_{{\rm z}, l} = \sqrt{(\theta_l^{\rm x})^2+(\theta_l^{\rm y})^2}$, and ${\phi}_l^{\rm z} = \sqrt{(\phi_l^{\rm x})^2+(\phi_l^{\rm y})^2}$ to obtain the  directions of departure and arrival,  i.e.  ${\boldsymbol \theta}_l = (\theta_l^{\rm x}, \theta_l^{\rm y}, \theta_l^{\rm z})$ and ${\boldsymbol \phi}_l = (\phi_l^{\rm x}, \phi_l^{\rm y}, \phi_l^{\rm z})$.

By applying the general expressions for the complexity of the MOMP algorithm in terms of the parameters of the sparse recovery problem  included in Fig.~\ref{fig:alg_diag}, we can obtain the computational complexity order of the classic OMP algorithm and the MOMP algorithm for the problem of channel estimation at mmWave as a function of the system parameters.
The complexity order can be described as a function of the parameter $K_{\rm res}$, defined as the dictionaries resolution factor defining their number of elements proportionally to their elements size as $N_k^{\rm a} = \lfloor K_{\rm res}N_k^{\rm s}\rfloor$.
It can be observed in Table~\ref{tab:complexity} that the complexity order of OMP depends on $K_{\rm res}^5$, while for MOMP depends only on $K_{\rm res}$.

\section{Localization}
\begin{figure}
\centering
\includegraphics[width=0.95\linewidth]{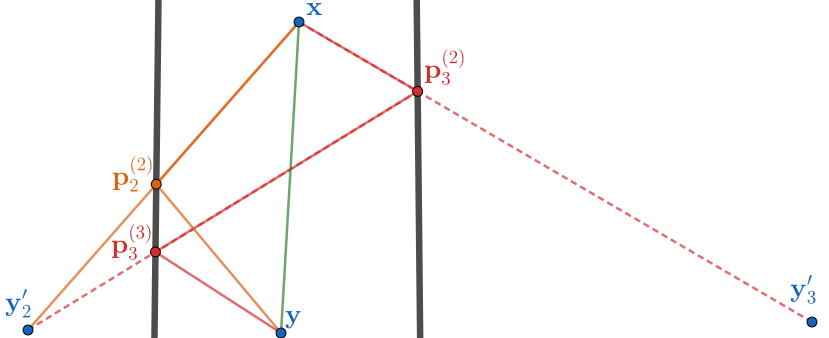}
\caption{Illustration of the idea of virtual anchor for localziation. In this image we can see both the path decomposition in multiple points and the virtual anchors created by specular reflections of the transmitter for the line of sight path $l=1$ (green), first order reflection $l=2$ (orange), and second order reflection $l=3$ (red).}
\label{fig:path_points}
\end{figure}

Given the geometric channel model, the channel paths can be understood as rays that begin at the transmitter and end at the receiver, after potentially bouncing in the objects in the environment. Therefore, it is possible to apply basic specular geometry tools to the problem of obtaining an estimation of the user position from the estimated channel paths.
To be more specific, each propagation path can be decomposed into a finite sequence of points ${{\bf p}_{l}^{(1)}, \ldots, {\bf p}_{l}^{(n_l)}}\subset\mathbb{R}^3$, such that the first and last points correspond to the receiver and the transmitter locations, that we will denote as ${\bf a}={\bf p}_{l}^{(1)}$ and ${\bf u}={\bf p}_{l}^{(n_1)}$ respectively, as depicted if Fig.~\ref{fig:path_points}.
Using this definition, ${\boldsymbol \theta}_l$ and ${\boldsymbol \phi}_l$ are directly related to the last two and first two points in the series as ${\boldsymbol \theta}_l=\frac{{\bf p}_{l}^{(2)}-{\bf p}_{l}^{(1)}}{\|{\bf p}_{l}^{(2)}-{\bf p}_{l}^{(1)}\|}$ and ${\boldsymbol \phi}_l=\frac{{\bf p}_{l}^{(n_l-1)}-{\bf p}_{l}^{(n_l)}}{\|{\bf p}_{l}^{(n_l-1)}-{\bf p}_{l}^{(n_l)}\|}$. Moreover, the delay is related to the path length as $\tau_l = \frac{1}{c}\sum_{n=1}^{n_l-1}\|{\bf p}_{l}^{(n+1)}-{\bf p}_{l}^{(n)}\|$, while $\alpha_l$ has a much more complicated expression depending on the reflective properties of the reflections.

We will exploit now the idea of virtual image as used in specular geometry. Thus, each path, as seen by each one of the devices (disregarding changes in $\alpha_l$), is equivalent to a line of sight path ($n_l=2$) when the other device's location is replaced with its specular reflection, commonly refered to as its virtual image.
In other words, for our localization problem, considering ${\bf u}'_l$ to be the virtual image of ${\bf u}$ through the optical changes that path $l$ experiments, we have that
\begin{equation}\label{eq:model_localization}
\begin{array}{rl}
{\boldsymbol \theta}_l = & \frac{{\bf u}'_l-{\bf a}}{\|{\bf u}'_l-{\bf a}\|}\\
\tau_l = & \frac{\|{\bf u}'_l-{\bf a}\|}{c}
\end{array}.
\end{equation}
These specular reflections are represented in Fig.~\ref{fig:path_points} for first and second order reflections.
The first order reflection virtual anchor is the direct specular reflection with the wall, while the second order reflection virtual anchor is a concatenation of specular reflections with both walls.

Considering the formulation in \eqref{eq:model_localization}, and having ${\bf a}$, ${\boldsymbol \theta}_l$ and $\tau_l-\tau_0$ for multiple paths, we can use this information to locate the user.
As a preliminary step,  and with the intention to extend this result to more complex scenarios, we will make the following assumptions: 1) the channel has a LoS component, so that the first computed path properties ${\boldsymbol \theta}_1$ and $\tau_1$ correspond to that of the line of sight path; 2)  walls are vertical; and 3)  floor/ceiling are horizontal. These assumptions will be exploited  to create a ranging system from the information about the parameters of the reflected and LoS paths.
The line of sight condition grants us with the equation
\begin{equation}\label{eq:los_loc}
{\bf u} = {\bf a}+\|{\bf u}-{\bf a}\|\frac{{\bf u}-{\bf a}}{\|{\bf u}-{\bf a}\|} = {\bf a}+c\tau_1{\boldsymbol \theta}_1,
\end{equation}
being ${\bf a}$ the receiver (access point) location and ${\bf u}$ the transmitter (user equipment) location, while the room geometry condition can be used to create a ranging system to solve $\tau_1$.

\subsection{Ranging}
First, we are going to use the fact that mmWave channels have poor reflective properties, thus virtually all relevant paths to the channel (the strongest ones) are going to either be line of sight ($n_l = 2$) or first order reflections $n_l = 3$.
The fact that the walls of the room are vertical tells us that any specular reflection of the user through a wall will be in the same horizontal plane. Therfore, the line of sight and reflections with walls travel the same vertical distance, that is $[{\bf u}]_{3} = [{\bf u}'_l]_{3}$, which can be expressed as 
\begin{equation}
\theta_l^{\rm z}\tau_l = \theta_1^{\rm z}\tau_1
\end{equation}
for any index $l$ corresponding to a wall reflection.
Analogously, because the floor and ceiling are horizontal, the specular reflection of the user through the floor/ceiling will be in the same vertical line. Therefore, the line of sight and reflections with floor/ceiling travel the same horizontal distance, that is $[{\bf u}]_{1} = [{\bf u}'_l]_{1}, [{\bf u}]_{2} = [{\bf u}'_l]_{2}$, which can be expressed as 
\begin{equation}
\sqrt{(\theta_l^{\rm x})^2+(\theta_l^{\rm y})^2}\tau_l = \sqrt{(\theta_1^{\rm x})^2+(\theta_1^{\rm y})^2}\tau_1
\end{equation}
for any index $l$ corresponding to a floor/ceiling reflection.

\begin{figure*}[!t]
\normalsize
\setcounter{MYtempeqncnt}{\value{equation}}
\setcounter{equation}{39}
\begin{equation}
\left\lbrace\begin{array}{ccc}
\theta_l^{\rm z}((\tau_l-\tau_0)+\tau_0) = \theta_1^{\rm z}((\tau_1-\tau_0)+\tau_0) & \text{if} & \text{wall reflection}\\
\sqrt{(\theta_l^{\rm x})^2+(\theta_l^{\rm y})^2}((\tau_l-\tau_0)+\tau_0) = \sqrt{(\theta_1^{\rm x})^2+(\theta_1^{\rm y})^2}((\tau_1-\tau_0)+\tau_0) & \text{if} & \text{floor/ceiling reflection}
\end{array}\right.
\label{eq:tauzeroest}
\end{equation}
\setcounter{equation}{\value{MYtempeqncnt}}
\hrulefill
\vspace*{4pt}
\end{figure*}

These geometric relations depend on the type of path. Therefore, a method to classify the different estimated paths has to be proposed. To this aim, we will make use of two properties: a) specular reflections on the horizontal plane (floor/ceiling) do not change the $x$-$y$ components of a point, and b) vertical reflections (walls) do not change the $z$ component. With this in mind, we consider four possible types of paths: line of sight (LoS), wall reflection, floor/ceiling reflection, or any other path
that will be labeled as spurious and will not be exploited for localization.

The inputs to our proposed path classification algorithm are the spherical coordinates of the different angles, computed as  $\theta^{\rm az} = \arg(\theta^{\rm x} + j\theta^{\rm y})$, $\theta^{\rm el} = \arcsin(\theta^{\rm z})$, $\phi^{\rm az} = \arg(\phi^{\rm x} + j\phi^{\rm y})$ and $\phi^{\rm el} = \arcsin(\phi^{\rm z})$.
Wall reflections and LoS paths satisfy $\theta_l^{\rm el} + \phi_l^{\rm el} = 0$, while floor/ceiling reflections satisfy $\theta_l^{\rm el} = \phi_l^{\rm el}$.
Additionally, floor/ceiling reflections and LoS paths arrival and departure azimuth angles are opposite, i.e. they are separated by $180^\circ$.
By defining the threshold values $r_{\rm az}, r_{\rm el}$, we can obtain the conditions
\begin{align}
|\sin(\theta^{\rm el}-\phi_l^{\rm el})| < r_{\rm el}\label{eq:condition_el_v},\\
|\sin(\theta^{\rm el}+\phi_l^{\rm el})| < r_{\rm el}\label{eq:condition_el_h},\\
\cos(\theta^{\rm az}-\phi_l^{\rm az}) < r_{\rm az}-1.\label{eq:condition_az}
\end{align}
A LoS path satiifies conditions \eqref{eq:condition_el_v} and \eqref{eq:condition_az}, floor/ceiling reflections satisfy \eqref{eq:condition_el_h} and \eqref{eq:condition_az}, while wall reflections only satisfy  \eqref{eq:condition_el_v}. Any other path will be classified as spurious.

Once we have classified the main reflections, we can solve the following system of linear equations defined in \eqref{eq:tauzeroest}.
The variables $(\tau_l-\tau_0)$ and ${\boldsymbol \theta}_l$ are extracted from the path decomposition, thus only $\tau_0$ remains to be estimated. Least squares estimation can be used to solve this overdetermined system.
Once $\tau_0$ has been estimated, we can plug it into  \eqref{eq:los_loc} together with the other known variables ${\bf a}$ and ${\boldsymbol \theta}_l$ to get the user location ${\bf u}$.

\section{Simulations}
\begin{figure}
\includegraphics[width=0.9\linewidth]{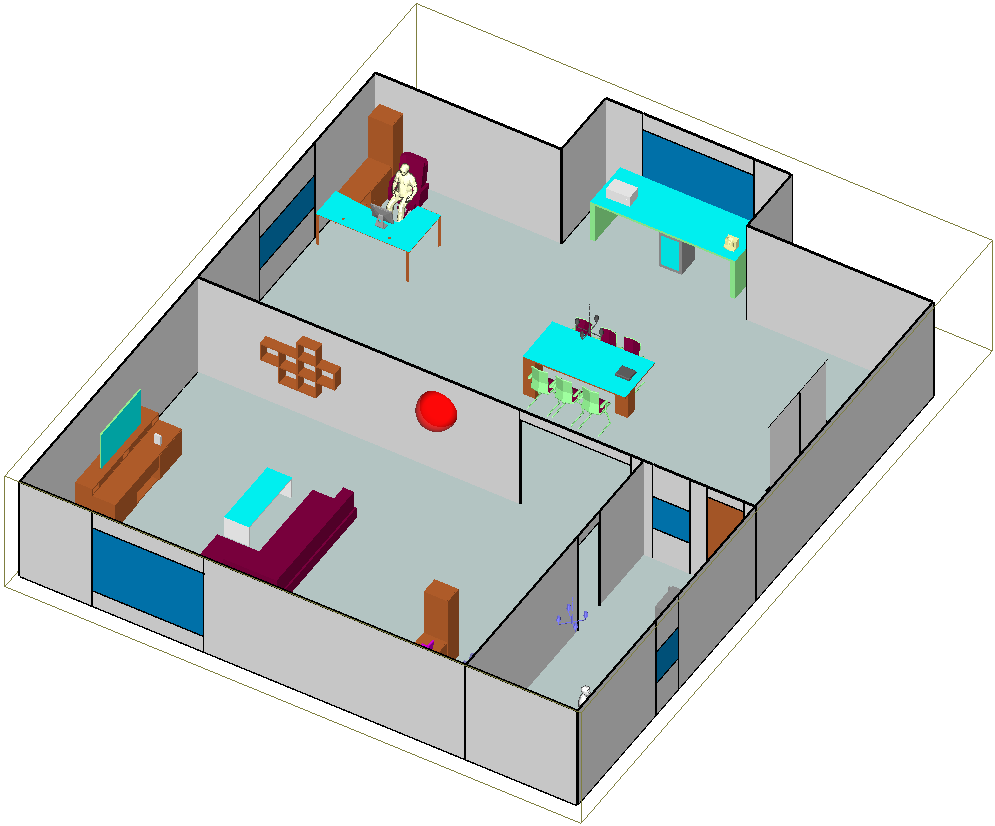}
\caption{3D view of the ray-tracing scenario. The two access points are located at both sides of the center wall, in the point represented by a red sphere.}
\label{fig:scenario}
\end{figure}

We consider a ray-tracing simulation of a home office implemented with Wireless InSite 3D wireless prediction software, depicted in Fig.~\ref{fig:scenario}.
We simulate the uplink of the mmWave communication system, with two access points at both sides of the center wall, in the point represented by a red sphere in Fig.~\ref{fig:scenario}. The scripts used to obtain the results provided in this section can be found in \cite{CodeMOMPPaper}.

For the antenna arrays,  we consider uniform planar arrays for the user and access points, of sizes $N_{\rm T}^{\rm x}\times N_{\rm T}^{\rm y}=4\times4$ and $N_{\rm R}^{\rm x}\times N_{\rm R}^{\rm y}=8\times8$, respectively. The access point exploits $M_{\rm R} = 8$ RF-chains, while the user employs a single one, i.e.  an analog beamforning architecture, $M_{\rm T} = 1$. The transmit power is initially set to $20{\rm dBm}$, a realistic value for indoor scenarios.
$\sigma^2$ is computed as the thermal noise corresponding to a room at $15^\circ{\rm C}$ when using a $2{\rm GHz}$ bandwidth, resulting in $-81{\rm dBm}$. The number of considered delay taps in the channel is 64, and the pulse shaping function is a sinc 
The training precoders and combiners in the initial experiments are built by following the procedure explained next. We start by creating the matrix resulting of the Kronecker product of the DFT matrices with sizes $N_{\rm R}^{\rm x}$ and $N_{\rm R}^{\rm y}$ for the device, and $N_{\rm T}^{\rm x}$ and $N_{\rm T}^{\rm y}$ for the access point.
Then, the matrix created at the device is split in columns, and the one created at the access point is split in blocks of $M_{\rm R}$ columns, thus creating $16$ different training analog precoders at the device and $8$ hybrid precoders of size $8$ at the access point.
The channel measurements contain all the different combinations of training precoders combiners, requiring $M=128$ training frames in total. The number of considered delay taps is 64 and the time response is a sinc function.
The pilot signal we are considering consists of $64$ ones, with 64 padded  zeros at the beginning of the pilot and 32 padded zeros at the end. 

For algorithm evaluation purposes, we consider the problem of localizing a user device at a height of $1.3{\rm m}$ at $218$ different locations following a path the expands for the main two rooms of the simulation environment.
We assume that the user connects to the access point with highest gain and compute the observation according to the adopted channel model.
We constrain the sparsity level to $N_{\rm p}=5$ hoping to get the LoS, $3$ wall reflections and the ceiling reflection.
This sparsity is enough to capture the maximum number the expected significant paths.
We don't want to increase this number to avoid the estimation of spurious paths that may affect negatively to the localization estimation.
To simplify the analysis, we make use of dictionaries of size proportional to the dimensions of their atoms, with a 
proportionally constant defined by the parameter $K_{\rm res}$, that is $N_k^{\rm a} = K_{\rm res}N_k^{\rm s}\quad\forall k\leq N_{\rm D}$.

\begin{figure}
\begin{tabular}{c}
\includegraphics[width=0.9\linewidth]{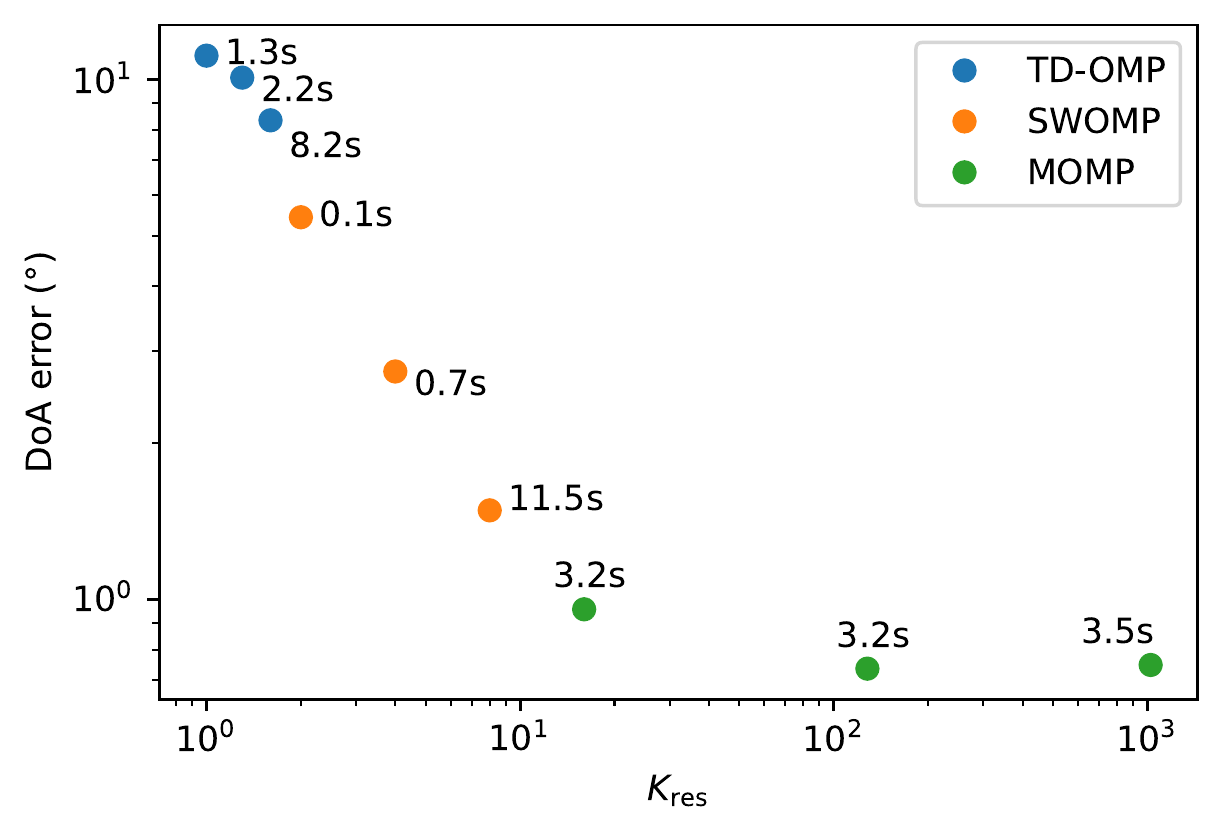}\\ (a) \\
\includegraphics[width=0.9\linewidth]{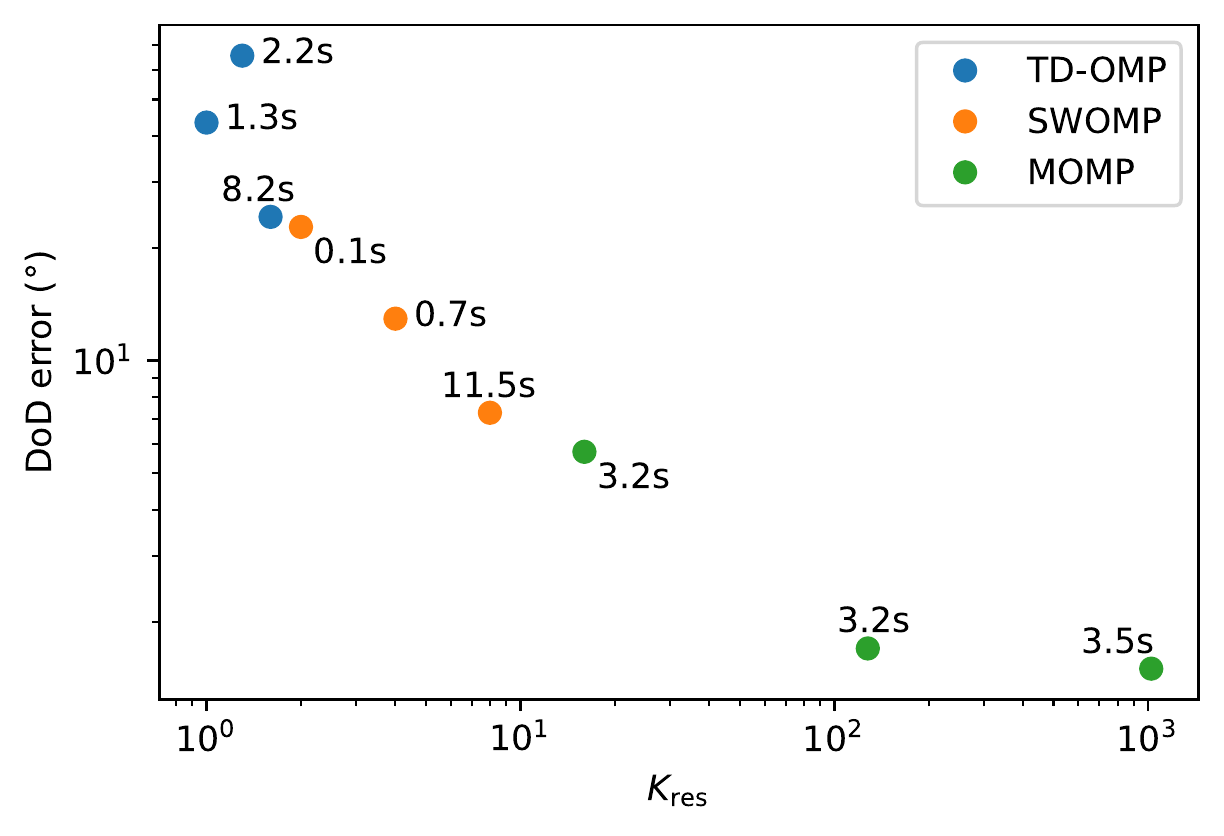}\\ (b)
\end{tabular}
\caption{Median main path angular error for different evaluations of OMP/MOMP. The text indicates the average running time.}
\label{fig:Ang_err}
\end{figure}

\begin{figure}
\begin{tabular}{c}
\includegraphics[width=0.9\linewidth]{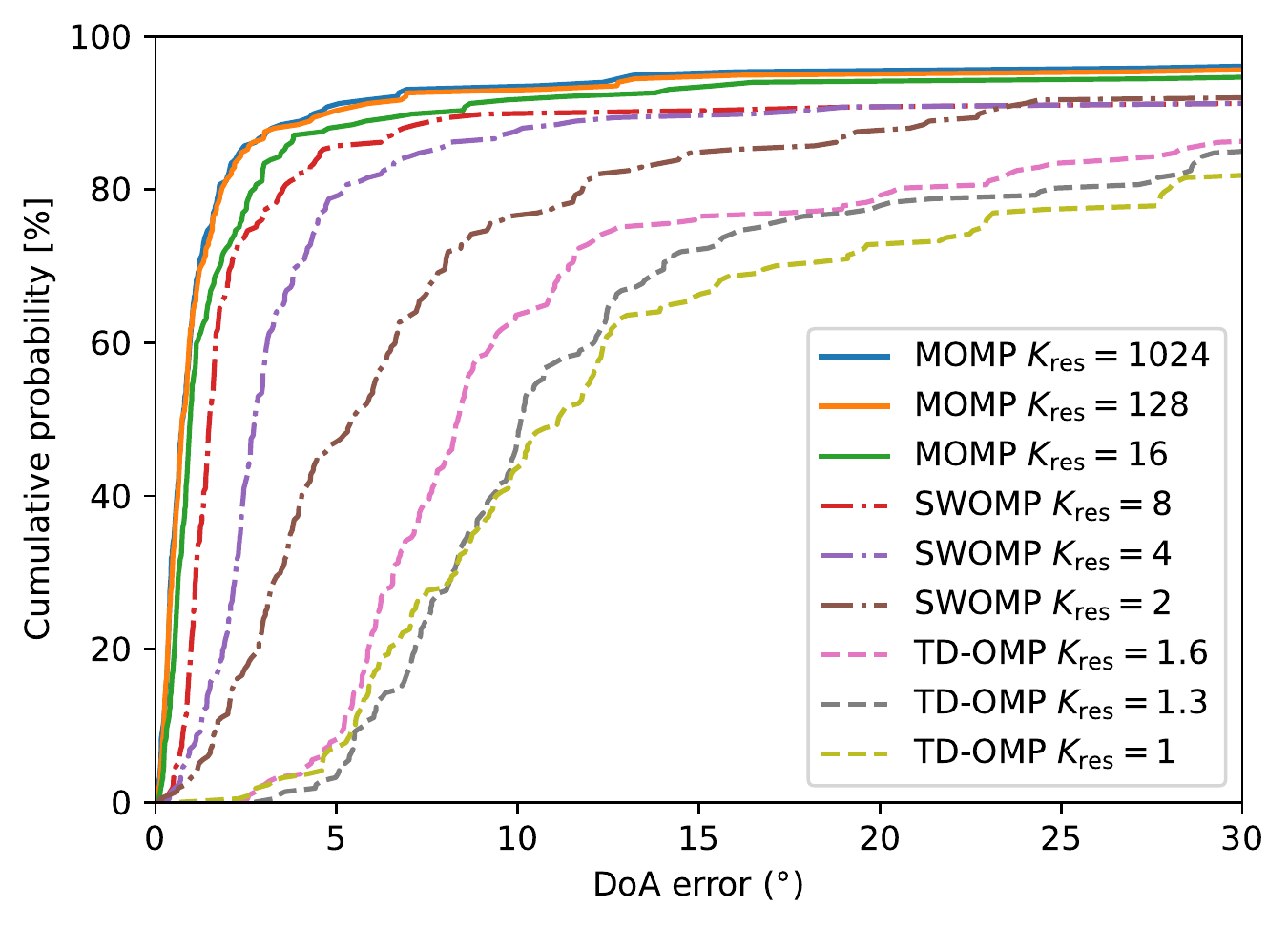}\\ (a)\\
\includegraphics[width=0.9\linewidth]{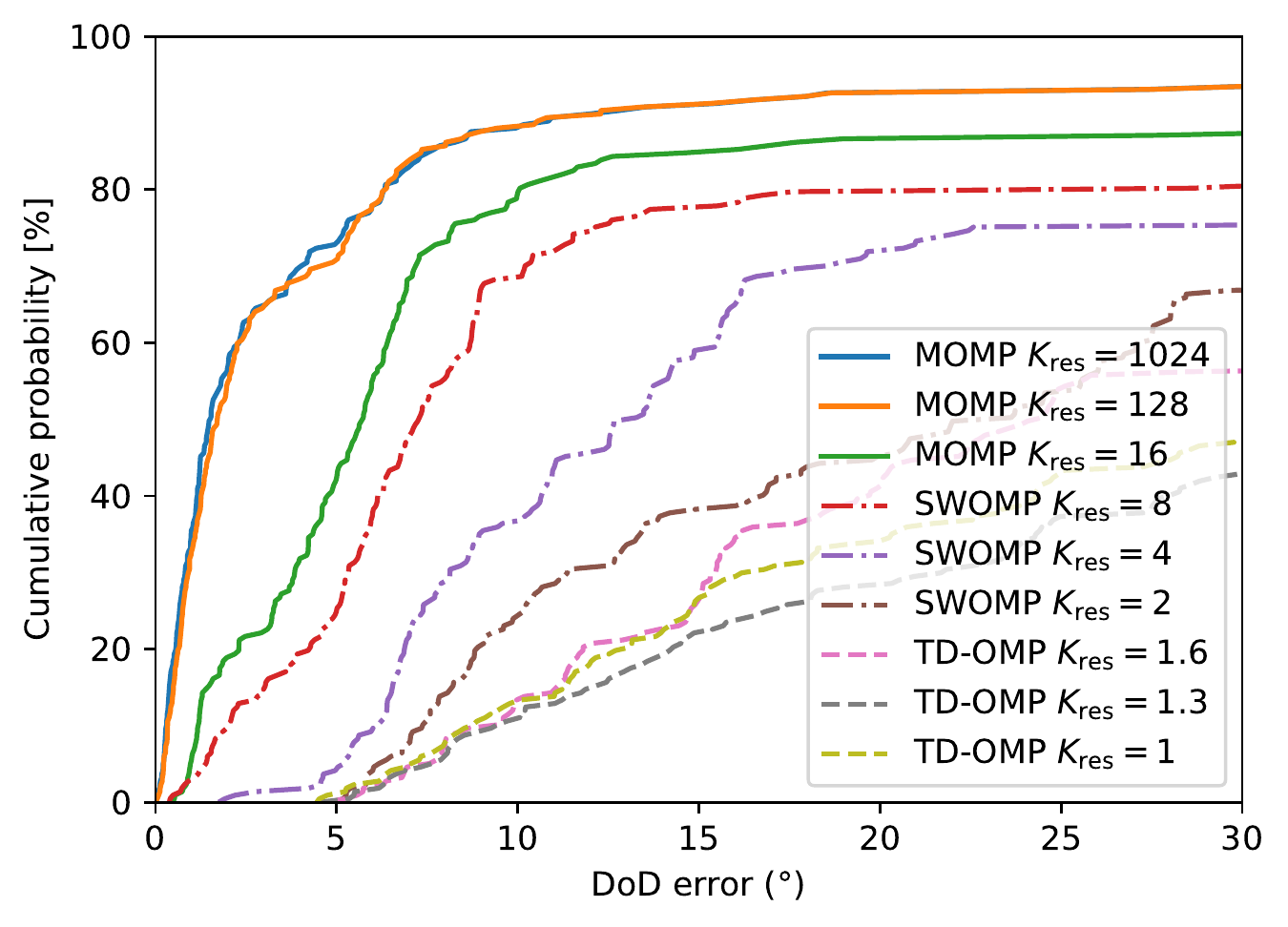}\\ (b)
\end{tabular}
\caption{Angular error CDFs for different evaluations of OMP/MOMP.}
\label{fig:DoA_DoD_err}
\end{figure}

In the first experiments, we compare the performance of MOMP to the one provided by the classic OMP algorithm when exploited by the time domain algorithm in \cite{Venugopal2017}, denoted as TD-OMP, and by the frequency domain algorithm in \cite{SWOMP2018}, denoted as SWOMP.
We consider values of $K_{\rm res}$ that fit the algorithm's computational complexity. For the sake of evaluating OMP without overflowing the memory, we consider small antenna arrays, setting $N_{\rm R}^{\rm x}=N_{\rm R}^{\rm y}=6$ and $N_{\rm T}^{\rm x}=N_{\rm T}^{\rm y}=3$.
We choose $K_{\rm res}\in[1, 1.2, 1.4]$ for OMP, since larger values will prevent OMP from running,  and $K_{\rm res}\in[16, 128, 1024]$ for MOMP.
The evaluated metric is the  median angular error for the main path, i.e. $\arccos(<\boldsymbol{\theta}_1, \hat{\boldsymbol{\theta}}_1>)$.
We can see the perfomance of both algorithms in Fig.~\ref{fig:Ang_err}, which shows results for the DoA in (a) and the AoD showing how MOMP outperforms the other two algorithms while keeping the computational complexity lower than the best performing cases of the other two algorithms.
Additionally, the cumulative distribution functions (CDFs) of these errors is also depicted in Fig.~\ref{fig:DoA_DoD_err} showing in more detail how MOMP is able to exploit larger dictionaries for more accurate estimations getting an estimation twice as accurate as SWOMP across most percentiles and leaving TD-OMP out of the competition.

\begin{figure}
\begin{tabular}{c}
\includegraphics[width=0.9\linewidth]{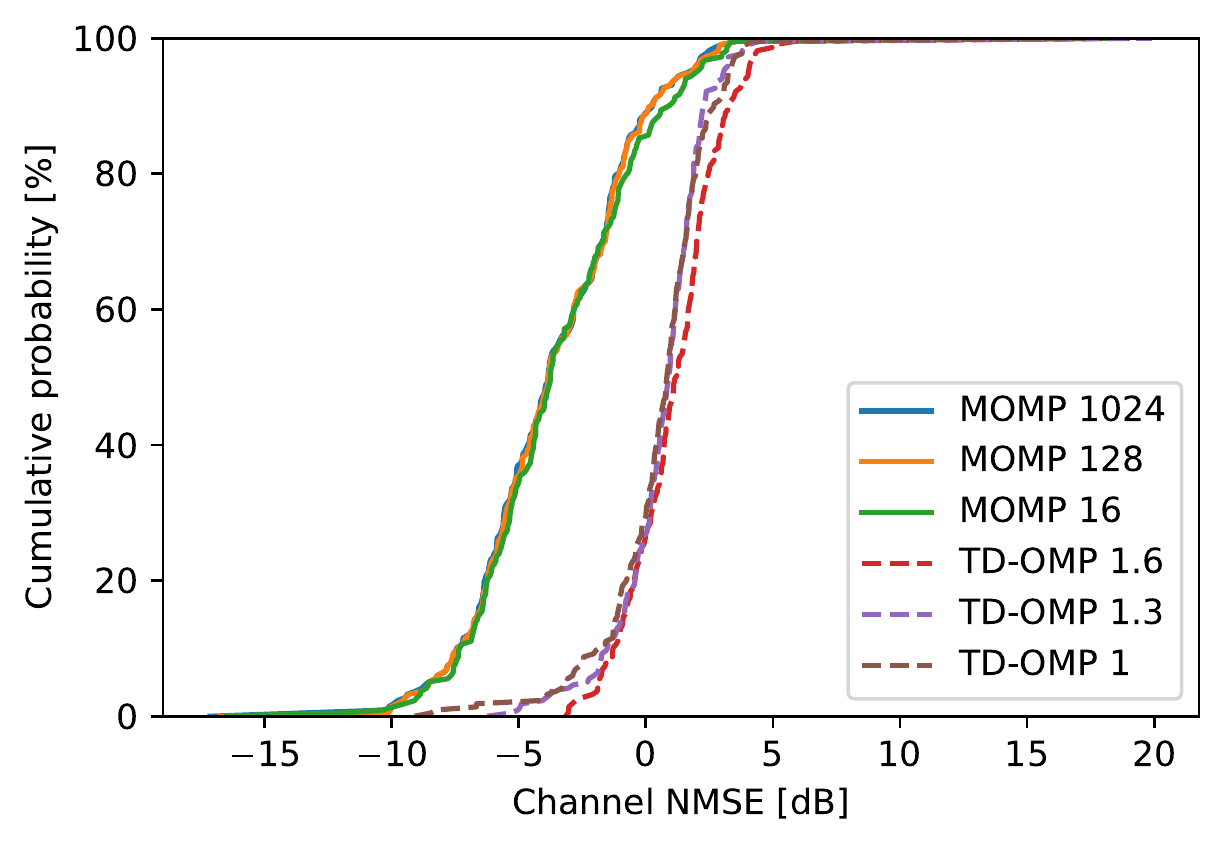}\\ (a)\\
\includegraphics[width=0.9\linewidth]{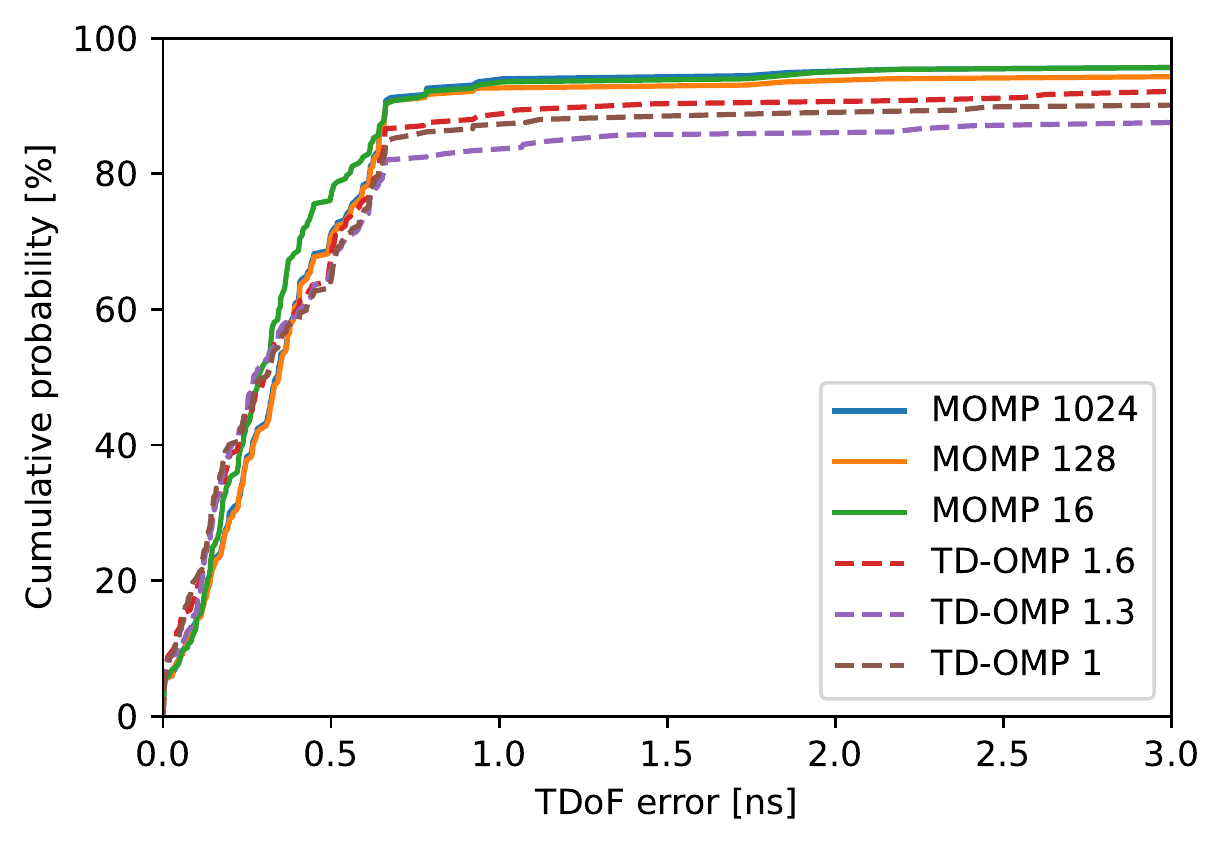}\\ (b)
\end{tabular}
\caption{(a): Normalized channel error CDF.
(b) Secondary path's angle difference of arrival error CDFs.
*Note: The number in the legend represent the value of $K_{\rm res}$.}
\label{fig:Ch_TDoF_err}
\end{figure}

Next, we analyze the normalized mean square error (NMSE) in the channel reconstruction, i.e.  $\frac{\sum_d\|\hat{\bf H}_d-{\bf H}_d\|^2}{\sum_d\|{\bf H}_d\|^2}$, and the secondary paths' delay error, i.e.  $\min_l(\tau_2-\hat{\tau}_l)$.
In order to evaluate these two metrics, to avoid in-determination caused by the unknown delay $\tau_0$ we assume the main path's delay $\tau_1$ to be known to compute $\tau_0$ and use it to compute each path's absolute delay as $\tau_l = (\tau_l - \tau_0) + \tau_0$.
This evaluation can be seen in the form of cumulative distribution functions in Fig.~\ref{fig:Ch_TDoF_err} which shows a $5{\rm dB}$ gap between the channel estimation from MOMP and the ones from SWOMP.
The time delay figure also shows a consistent uniform time delay error ranged from $0$ to $0.7{\rm ms}$ showing a not detection probability twice as big for SWOMP than MOMP.

\begin{figure}
\begin{tabular}{c}
\includegraphics[width=0.9\linewidth]{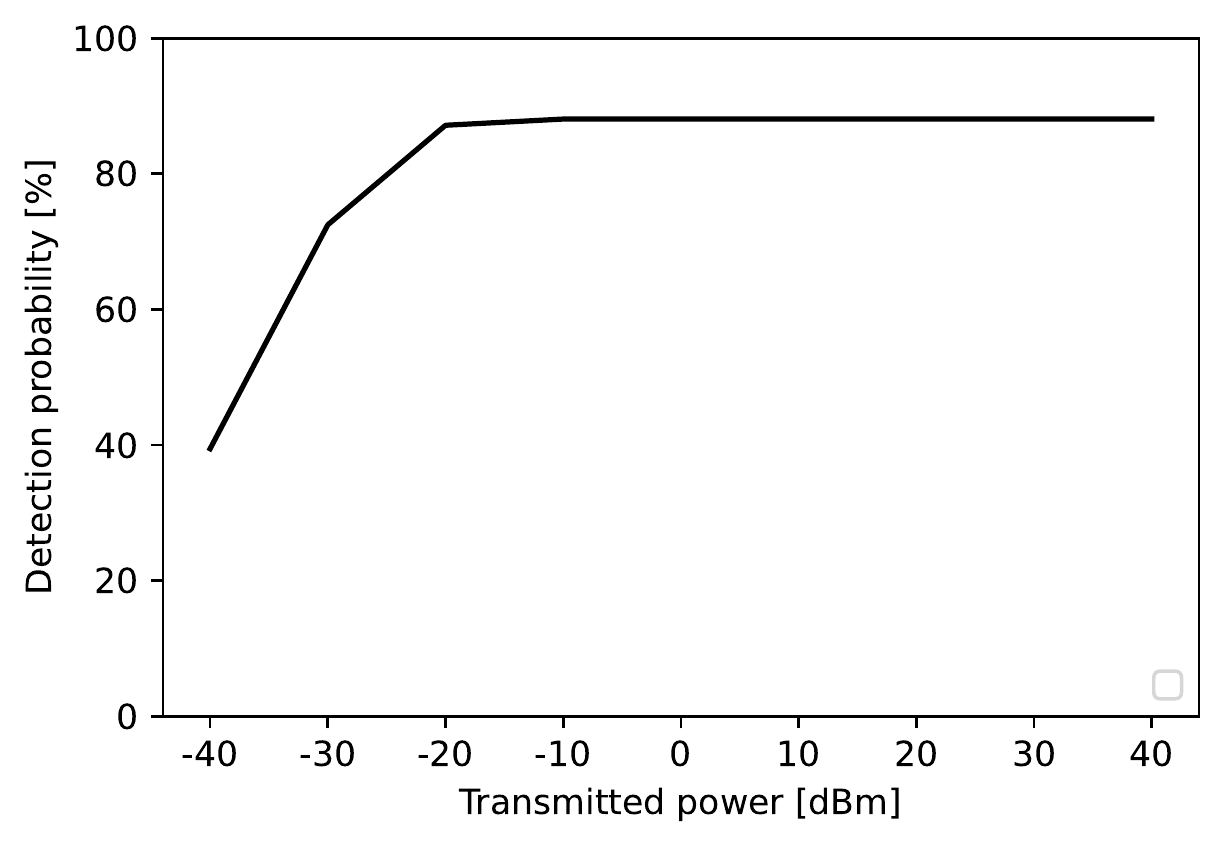}\\ (a)\\
\includegraphics[width=0.9\linewidth]{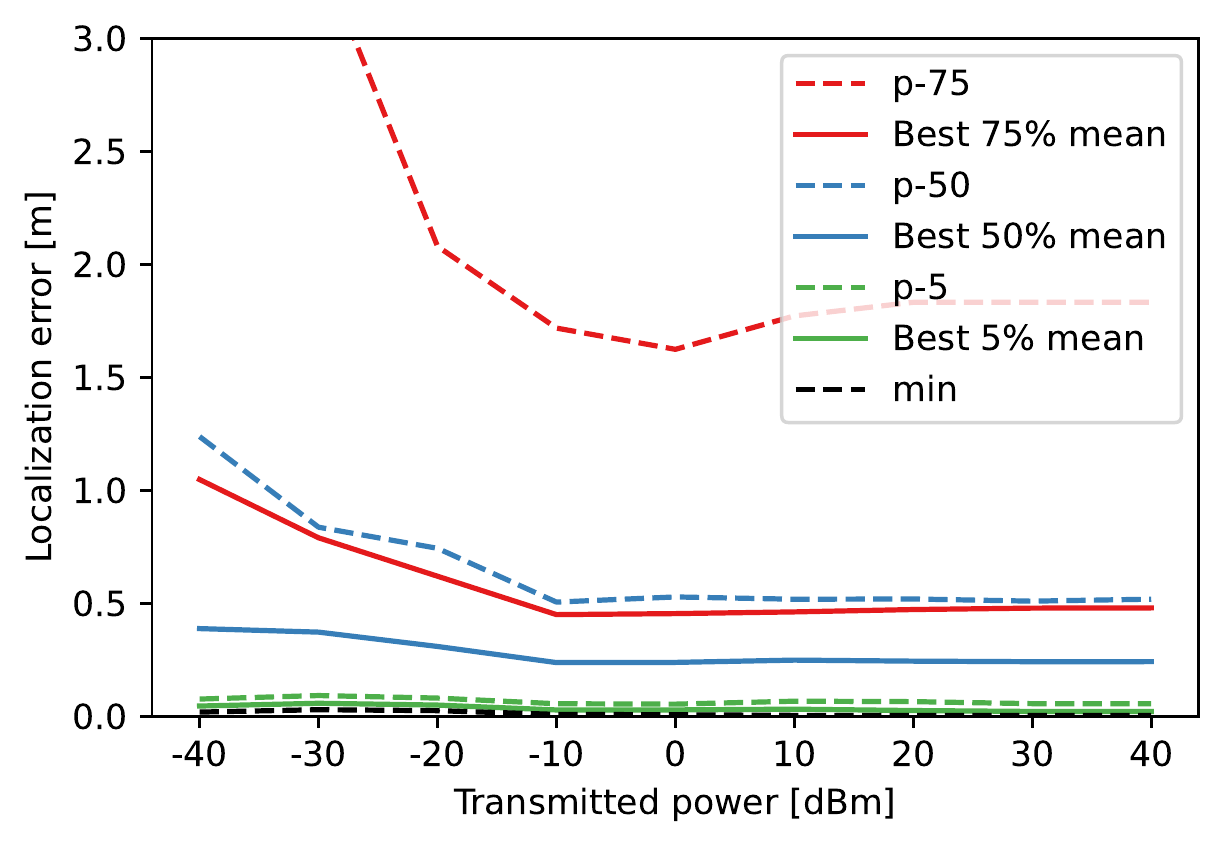}\\ (b)
\end{tabular}
\caption{(a): Localization detection for different transmitter power.
(b): Localization error for different transmitter power, percentiles and average of best $X\%$ cases.}
\label{fig:LocErr_Pt}
\end{figure}

The next simulations evaluate the localization accuracy when varying the transmitted power.
Note that, since this localization algorithm depends on an undetermined system of equations, the algorithm may fail to produce an answer if not enough paths get classified in the desired categories.
This translates into not only getting an error but also a probability of detection specially when the number of paths is low like in a NLoS scenario case.
Both detection and error are reflected in Fig.~\ref{fig:LocErr_Pt} in which we see how the localization accuracy doubles quasi-linearly from $-40{\rm dB}$ to $-10{\rm dB}$ and then saturates while under $-20{\rm db}$ every decrease of $-10{\rm dB}$ in SNR results in about a $20\%$ increase of undetected cases.

Finally, we study the impact of the number of training frames in the localization accuracy and the improvement that can be achieved by using a hybrid MIMO  architecture at the user side instead of analog. 
As training combiners at the access point we use the Kronecker product of to $X_{\rm R}$ sector beam patterns designed for linear antennas of $N_{\rm R}^{\rm x} = N_{\rm R}^{\rm y}$ elements, following the design in \cite{JoanLight}, with equally spaced steering directions and a beam width of $\frac{2\pi}{X_{\rm R}}$.
Analogously,  for the training precoders, we consider the Kronecker product of $X_{\rm T}$ sector beam patterns designed for linear antennas of $N_{\rm T}^{\rm x} = N_{\rm T}^{\rm y}$ elements, following again the design in \cite{JoanLight}, with equally spaced steering directions and a beam-width of $\frac{2\pi}{X_{\rm T}}$.
For the hybrid architecture at the user side we choose $N_{\rm T}^{\rm RF} = 4$, and we use  a pilot signal ${\bf S}$ defined by the first $N_{\rm T}^{\rm RF} = 4$ rows of a 64 element Hadamard matrix, normallized by a factor $\frac{1}{\sqrt{N_{\rm T}^{\rm RF}}}$, with zero padding as described for the analog architecture. This leads to a total number of  $(X_{\rm R}X_{\rm T})^2/(N_{\rm R}^{\rm RF}N_{\rm T}^{\rm RF})$ training frames.
We select the number of pairs $(X_{\rm R}, X_{\rm T})\in\{(2, 4), (2, 8), (4, 8), (4, 16), (8, 16)\}$.
\begin{figure}
\includegraphics[width=0.9\linewidth]{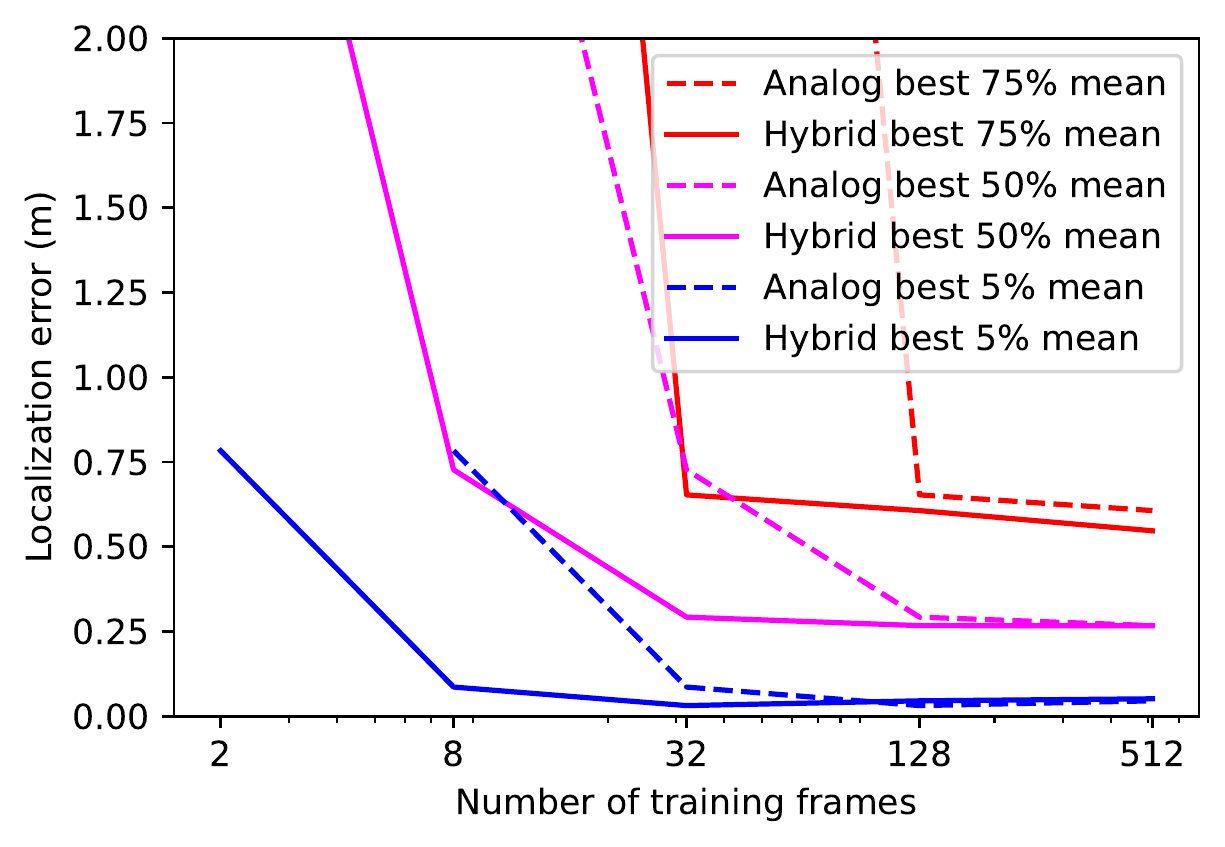}
\caption{Average localization error of the best $X\%$ cases for different number of training frames with different UE antenna architectures.}
\label{fig:LocErr_TF}
\end{figure}
Fig.~\ref{fig:LocErr_TF} shows a comparison of the localization performance gains that come from increasing the number of training frames and/or making use of a hybrid architecture at the user with hybrid architecture outperforming Analog architecture and both saturating to a constant accuracy at $128$ frames for the Analog architecture and $32$ frames for the Hybrid architecture.

\bibliographystyle{IEEEtran}  
\bibliography{refs,refs_kron}     


\end{document}